\documentclass[11pt]{article}
\usepackage[margin=1in]{geometry} 
\usepackage{textcomp}
\usepackage{gensymb}
\usepackage{graphicx}
\usepackage{amssymb,amsmath}
\usepackage{amstext,amsthm, amssymb,amsfonts}
\usepackage{epstopdf}
\usepackage[usenames,dvipsnames]{xcolor}
\usepackage{multirow}
\usepackage{placeins}
\usepackage[linesnumbered,ruled,vlined]{algorithm2e}
\usepackage{complexity}
\usepackage{soul}
\usepackage{amssymb,amsfonts,amsmath,amsthm}
\newtheorem{makeproposition}{Proposition} 
\newtheorem{makelemma}{Lemma} 

\newcommand{\xhdr}[1]{\paragraph*{\bf #1}}

 \newcommand{\abar}[2]{{f^{#1}_{#2}}}
 \newcommand{\bbar}[2]{{g^{#1}_{#2}}}
\def\bra{\left\langle}
\def\ket{\right\rangle}
\def\inpathsVar{\widehat{A}} 
\def\outpathsVar{\widehat{B}} 
\def\inprobrandVar{\widehat{a}} 
\def\outprobrandVar{\widehat{b}} 
\def\pin{p_{in}}
\def\pout{p_{out}}
\def\tif{\text{ if }}

\def\telse{\text{ else}}

\title{Block Models and Personalized PageRank}

\author{
  Isabel Kloumann\\
  Center for Applied Mathematics\\
  Cornell University\\
  \texttt{imk36@cornell.edu}
  \and
  Johan Ugander\\
  Management Science \& Engineering\\
  Stanford University\\
  \texttt{jugander@stanford.edu}
  \and
  Jon Kleinberg\\
  Department of Computer Science\\
  Cornell University \\
  \texttt{kleinber@cs.cornell.edu}
}

\begin{document}

\maketitle
\begin{abstract}
Methods for ranking the importance of nodes in a network have a rich history 
in machine learning and across domains that analyze structured data. 
Recent work has evaluated these methods though the \textit{seed set expansion
problem}: given a subset $S$ of nodes from a community of interest
in an underlying graph, can we reliably identify the rest of the
community? We start from the observation that the most widely used
techniques for this problem, personalized PageRank and heat kernel
methods, operate in the space of \textit{landing probabilities} of a
random walk rooted at the seed set, ranking nodes according to
weighted sums of landing probabilities of different length walks.
Both schemes, however, lack an a priori relationship to
the seed set objective. In this work we develop a principled framework
for evaluating ranking methods by studying seed set expansion applied
to the stochastic block model. We derive the optimal gradient for
separating the landing probabilities of two classes in a
stochastic block model, and find, surprisingly, that 
under reasonable assumptions the gradient is
asymptotically equivalent to personalized PageRank for a specific
choice of the PageRank parameter $\alpha$ that depends on
the block model parameters. This connection provides a novel
formal motivation for the success of personalized PageRank in seed set
expansion and node ranking generally. We use this connection to
propose more advanced techniques incorporating higher moments of
landing probabilities; our advanced methods exhibit
greatly improved performance despite being simple linear
classification rules, and are even competitive with belief
propagation.
\end{abstract}

\section{Introduction}

The challenge of contextually ranking nodes in a network
has emerged as a problem of canonical significance
in many 
domains, 
with a particularly rich history of study in social and information networks \cite{page1998pagerank,kleinberg1998authoritative,gupta2013wtf,gleich2015pagerank}.
An active line of recent work has focused on the problem of 
{\em seed set expansion} in networks 
\cite{andersen2006communities,
bagrow2008evaluating,
mehler2009expanding,
riedy2011detecting,
whang2013overlapping,
yang2012definingevaluating,kloumann2014community},
a fundamental version of node ranking
with the following natural definition.

In the seed set expansion problem,
we are given a graph $G$ representing some form of social
or information network, and there is a hidden community of
interest that we would like to find, corresponding to an internally
well-connected set of nodes.
We know a small subset $S$ of the nodes in this community,
and from this ``seed set'' $S$ we would like to expand outward
to find the rest of the community --- 
by ordering the rest of the nodes outside $S$
according to some ranking criterion, and proposing nodes
in this order as additional members of the community.
This problem arises in a wide range of domains, 
including settings where we are trying 
to find web pages that are related to a set of examples,
to identify a
social group from a set of sample members provided
by a domain expert, 
or to help a user automatically populate a group they
are defining in an online social-networking application.

A recent focus in the work on this problem has been the power of
approaches based on random-walk methods, including versions of
{\em personalized PageRank} \cite{haveliwala2002topic,jeh2003scaling,kloumann2014community} 
and physical analogues based on the
heat equation for graphs \cite{chung2007heat,kloster2014heat}.
These techniques can be viewed as operating on the following quantities:
for each node $v$ in the graph, and each number of steps $k$,
we let $r_k^v$ denote the probability that a random walk on the
graph ends up at $v$ after exactly $k$ steps, starting from
a particular seed node in $S$ (or a node chosen uniformly at
random from $S$).
Methods based on PageRank and heat kernels then combine these
values $\{r_k^v\}$ using particular functional forms as 
{\em discriminant functions}---a phase coined by Fisher to describe
functions for classification \cite{fisher1936use}---that 
produce a ``score'' for each node $v$ 
with the structure $score(v) = \sum_{k = 1}^\infty w_k r_k^v$
for coefficients $\{w_k\}$. The seed
set can be expanded by considering nodes in 
decreasing order by score
\cite{kloster2014heat,kloumann2014community}.
Geometrically, these rankings amount to sweeps through the space
of landing probabilities with hyperplanes normal to some vector,
where personalized PageRank and the heat kernel method
correspond to different choices of normal vectors. 

These methods are elegant in their formulation and have also shown to be
both quite powerful and scalable \cite{kamvar2003exploiting,bahmani2010fast,kloster2014heat}.
At the same time, their success has left open a number of
very basic questions.  In particular, if we think of the 
{\em landing probabilities}
$\{r_k^v\}$ over nodes $v$ and steps $k$ as providing us with a rich
set of features relevant to membership in the community of interest,
then it becomes clear that personalized PageRank and heat kernel 
formulations are simply specific, and apparently arbitrary,
ways of combining these features using hand-constructed weight coefficients
$\{w_k\}$. 

Motivations for the specific form of these two scores have come from
several domains. These include
the {\em random surfer model} for PageRank \cite{page1998pagerank} 
consisting of a mixture of random-walk steps and random jumps, as well 
as results connecting both PageRank and heat kernel quantities to
bounds on sparse cuts \cite{andersen2006local,chung2009local}
and regularized solutions to the min-cut problem \cite{gleich2014anti}.
Even here, however, there has not been an argument that any of these
measures are optimally combining the random walk landing probabilities under
a specific objective, nor has there been a direct connection between
any of these measures and the problem that seed expansion
seeks to solve.

Is there a principled reason why the expressions for PageRank or the heat
kernel represent the ``right'' way to combine the information coming 
from random walks, or could there be better approaches? And is there a formal
framework available for deriving or at least motivating effective
ways of combining random walk probabilities?  
Given the diverse and 
important applications where PageRank and heat kernel methods 
have seen successes,
we consider a broader examination of the space of 
methods for combining available random walk 
information, appreciating that the approaches in existing work are
simply particular points in that space.

The key observation we pursue in this work is that
a basic model of separable structure in graphs
known as the stochastic block model \cite{holland1983stochastic}
can be employed to model the presence of a seed set in the graph, 
allowing us to derive principled methods for ranking nodes in 
the space of landing probabilities.
We focus our attention on a two-block stochastic block model,
where one block of nodes corresponds to the community of a labelled seed set,
while the other block of nodes corresponds to its complement, 
the remainder of the graph.
In this setting the problem 
of finding the hidden community of interest has a correct answer
with respect to an underlying graph generation process, and hence
methods for combining landing probabilities of random walks
can be evaluated according to their ability to find this answer.

For this two-block stochastic block model we first
derive the centroids, for each block, in the space
of landing probabilities. Studying this space, we make the 
surprising observation that the optimal hyperplane for 
performing a linear sweep between the two centroids
is asymptotically concentrated for large graphs
on the weights of personalized PageRank,
for a specific choice of the PageRank parameter corresponding to parameters
of the stochastic block model. 
This connection between personalized PageRank and stochastic
block models is a novel bridge between two otherwise disconnected
literatures, and gives a strong motivation for using personalized PageRank 
in contextual ranking and seed set expansion problems.

Beyond simple linear discriminant rules,
we observe block models
can be used to propose more advanced scoring
methods in the space of landing probabilities,
and our analysis points to important ways in which personalized
PageRank can be strengthened. Although its geometric orientation is optimal
with regards to the landing probability centroids, 
personalized PageRank does not account for 
the variance or covariance of these landing probabilities, e.g. how the 2-hop landing
probabilities from a given seed correlate with the 3-hop landing probabilities 
from that seed.
We derive weights that correctly incorporate
these variances and covariances 
and we show that relative to the stochastic block model benchmark,
this new family of measures significantly outperforms personalized PageRank.

\section{Discriminant Functions for Stochastic Block Models}

The {\em stochastic block model} (SBM) \cite{holland1983stochastic}, also known as the planted partition model \cite{condon2001algorithms,dyer1989solution}, is a distribution over 
graphs that generalizes the Erd\H{o}s-R\'{e}nyi random graph model
 $G(n,p)$ \cite{erdos1959} to include a planted block structure. 
It can be described in terms of the following process for constructing
a random graph $G = (V,E)$.
There is a partition of the nodes into $C$ disjoint sets (blocks)
$V_1,...,V_C$, where $|V_i| = n_i = \pi_i n $ and $\sum_{i=1}^C\pi_i=1$,
together with a a $C\times C$ matrix $P$ whose entries are
in $[0,1]$.
The entry $p_{ij}$ of matrix $P$
specifies the probability of a node in $V_i$ being connected
to a node in $V_j$:
$\Pr( (u,v) \in E | u \in V_i, v \in V_j) = p_{ij}$. 
Each edge is formed independently of all others.

A stochastic block model is thus completely described by the parameters $n$, 
$\pi=(\pi_1,...,\pi_C)$, and $P$, and we let $G(n,\pi,P)$ denote the distribution over
graphs with the given parameters.
We allow self-loops and derive results for both directed and undirected graphs, 
where the latter case constrains $P$ to be symmetric.
The Erd\H{o}s-R\'{e}nyi random graph model
$G(n,p)$, with $n$ and $p$ scalar, is an undirected one-block special case.

We will first place particular focus on two-block models ($C=2$), where the block 
$V_a$ denotes the community of the seed set (the {\em seed community}) 
and the block $V_b$ denotes the remainder of the graph. 
This two-block special case is sometimes 
also known as the {\em affiliation model} \cite{frank1982cluster}.
A key point is that the nodes in $V_a$ and $V_b$ will have different
landing probabilities, suggesting that discriminant functions in 
the space of landing probabilities can be used to classify node
membership in $V_a$.
We then generalize this approach to the case of $C > 2$ blocks,
where we seek a binary classification between nodes
in two aggregate classes, ``in'' and ``out'', 
now possibly composed of several heterogeneous blocks,
$V_a = \cup_{i\in S} V_i$ for $S\subset\{1,\dots,C\}$, 
and $V_b = \cup_{i\in T} V_i$ for $T = \{1,\dots,C\}\setminus S$.

To perform our classifications, 
we will focus on two particular classes of discriminant functions: 
geometric discriminant functions and Fisherian discriminant functions.
{\em Geometric discriminant functions} 
perform a linear sweep through the 
feature space from one centroid $a$ to another centroid $b$,  
$f(r)=w^T r$ for $w=(a-b)$,
scoring points based on their inner product with the vector
connecting the two centroids. 
This score will increase as 
one moves from $b$ to $a$ in the space.
{\em Fisherian discriminant functions} 
employ a descriptive model 
where the two classes of points  
are described by their first two moments using
 multivariate Gaussians $N(a, \Sigma_a)$ and $N(b,\Sigma_b)$
in the feature space, scoring 
points based on their relative probabilities
of belonging to the two Gaussians. The special case
of $\Sigma_a = \Sigma_b = I$ is equivalent
to a geometric discriminant function, but in general
the Fisherian approach can account for
heterogeneous variance and 
covariance across and between features to make a more
principled discrimination. 

\subsection*{Geometric discriminant functions}
Recall that $r_k^v$, the $k$-step landing probability of a node $v$ given a
seed node in an underlying graph, is the probability that a random
walk beginning at that seed node will be at $v$ after exactly $k$ steps.
We map each node $v$ to a vector of its first $K$
landing probabilities $(r_1^v, r_2^v, \ldots, r_K^v)$, and ask
what these vectors look like in graphs generated by a stochastic block model.

Let $(a_1,...,a_K)$ denote the centroid of the landing probabilities
for nodes in the in-class, and let
$(b_1,...,b_K)$ denote the centroid of the landing probabilities
for nodes in the out-class.  
The geometric discriminant function $f(r^v) = (a-b)^T r^v$
will then rank each node $v \in V$ 
based on the node's
vector of landing probabilities
$(r^v_1,\ldots,r^v_K) \in [0,1]^K$. 
In this notation personalized PageRank 
assigns scores according to the infinite sum 
$\sum_{k = 1}^\infty \alpha^k r_k^v$, for a parameter $\alpha \in (-1,1)$, and 
the heat kernel method assigns scores by 
$\sum_{k = 1}^\infty \frac{e^{-t} t^k}{k!} r_k^v$
for a parameter $t>0$. Truncating these methods to
a finite walk length $K$, both methods then amount to
linear discriminant functions for particular weight vectors
$w_{PPR} (\alpha) = (\alpha, \alpha^2, ..., \alpha^K)$ 
and $w_{HK}(t) = (e^{-1} t, \frac{e^{-2} t^2}{2}, ..., \frac{e^{-t} t^K}{K!})$. 
Note that the PageRank parameter $\alpha$ is often
interpreted as the ``teleportation'' probability of a teleporting random walk,
requiring $\alpha$ be non-negative, but under the above interpretation
the personalized PageRank is well-defined for $-1 < \alpha < 0$ as well.

\def\pin{p_{in}}
\def\pout{p_{out}}
\subsection{PageRank and SBMs: Two Identical Communities}
We now establish an asymptotic equivalence between
personalized PageRank and geometric classification of stochastic block models 
in the space of landing probabilities, the main theoretical result
of our work. We begin by stating and proving our results
for the special case of an SBM with two identical communities.
We then state the more general connection between personalized PageRank and
the geometric discriminant weight vector for SBMs with $C$ non-identical blocks.
A proof of the more general result is given in the appendix.

\vspace{2mm}
\begin{makeproposition}
\label{p:1}
Let $G_n$ be an $n$-node graph generated from a two-block stochastic block model 
$G(n,\pi,P)$ with equally sized communities ($\pi_1=\pi_2=1/2$), $N= n/2$, and
with $p_{11} = p_{22} = p_{in}>0$ and $p_{12} = p_{21} = p_{out}>0$.
Let $\hat w = \hat a - \hat b$ be the geometric discriminant weight vector in the space
of landing probabilities ($1$-step through $K$-step, $K$ fixed) 
between the empirical block centroids $(\hat a_1, \dots, \hat a_K)$ and $(\hat b_1, \dots, \hat b_K)$.

For any $\epsilon, \delta>0$, there exists an $n$ sufficiently large such
that 
$
||N \hat w - N \Psi ||_1 \le \epsilon
$
with probability at least $1-\delta$, where:

(a) $\Psi = \Psi(p_{in}, p_{out},n)$
is a vector with the $k^{th}$ coordinate specified by 
$\Psi_k= \frac{1}{N} (\frac{A_k - B_k}{A_k + B_k} )$ 
where $A_k$ and $B_k$ are solutions to the linear homogeneous matrix recurrence relation:
\begin{eqnarray}
  \left\{
\begin{aligned}
A_k &= N (\pin A_{k-1} + \pout B_{k-1})\\
B_k &= N (\pout A_{k-1} + \pin B_{k-1}),
\end{aligned}
\right.
\label{eqn:recurstate}
\end{eqnarray}
with initial conditions $A_0=1$, $B_0=0$.

(b) this recurrence relation can be solved exactly, leading to 
a closed-form expression for $\Psi$:
$$\Psi_k = \dfrac{1}{N} \left(\dfrac{p_{in} - p_{out}}{p_{in} + p_{out}}\right)^k,$$
and thus the geometric discriminant weight vector $\hat w$
is asymptotically equivalent to $w_{PPR}(\alpha)$
for $\alpha = \dfrac{\pin - \pout}{\pin + \pout}$.
\end{makeproposition}

The above proposition 
relies on the following lemma.

\begin{makelemma}[\bf Concentration for $C=2$ identical blocks]
\label{thelemma}
Let $G_n$ be an $n$-node graph generated from a two-block stochastic block model 
$G(n,\pi,P)$ with equally sized communities ($\pi_1=\pi_2=1/2$), $N= n/2$, and
$P$ fixed with $p_{ij}>0$, $\forall i, j$, where the first block is designated the seed block.

For any $\epsilon, \delta>0$,
there is an $n$ sufficiently large
such that the random landing probabilities 
$(\hat a_1,....,\hat a_K)$ 
and $(\hat b_1,...,\hat b_K)$
for a uniform random walk on $G_n$ starting in the seed block
satisfy the following conditions with probability at least $1-\delta$ for 
all $k>0$:
\begin{align}
N \hat a_k 
\in &
  \left[(1 - \epsilon) 
  \frac{A_k}{A_k + B_k}
,
  (1 + \epsilon) 
  \frac{A_k}{A_k + B_k}
\right]
\text{ and }
\\
N \hat b_k
\in &
  \left[(1 - \epsilon) 
  \frac{B_k}{A_k + B_k}
  ,
  (1 + \epsilon) 
  \frac{B_k}{A_k + B_k}
  \right],
\label{eqn:containstate2}
\end{align}
where $A_k$, $B_k$ are the solutions to the
matrix recurrence relation
\begin{eqnarray*}
  \left\{
\begin{aligned}
A_k &= N (\pin A_{k-1} + \pout A_{k-1})\\
B_k &= N (\pout B_{k-1} + \pin B_{k-1}),
\end{aligned}
\right.
\end{eqnarray*}
with $A_0=1$, $B_0=0$.
\end{makelemma}

A proof of the lemma is given in 
in the appendix. 
With the lemma as given we now prove Proposition~\ref{p:1}.

\begin{proof}[Proof (of Proposition 1)]
First we will use the lemma to show that the coordinates of 
the weight vector $\hat w=\hat a - \hat b$ are concentrated as specified. 
From Lemma 1 we have that for any $\epsilon_1>0, \delta>0$ there exists an $n$ sufficiently large such that 
\begin{align}
(1 - \epsilon_1) \frac{A_{k}}{A_{k} + B_{k}}
<&\ N \hat a_{k} <
(1 + \epsilon_1) \frac{A_{k}}{A_{k} + B_{k}}\\
(1 - \epsilon_1) \frac{B_{k}}{A_{k} + B_{k}}
<&\ N \hat b_{k} <
(1 + \epsilon_1) \frac{B_{k}}{A_{k} + B_{k}}
\end{align}
with probability at least $1-\delta$.

As a result, whenever this containment holds we have that 
\begin{align}
\dfrac{A_k - B_k}{A_k + B_k} - \epsilon_1 < N(\hat a_k - \hat b_k) < \dfrac{A_k - B_k}{A_k + B_k} + \epsilon_1.
\end{align}
This expression can be simplified by recasting the 
central quantities in terms of the geometric discriminant weight vector $\hat w = \hat a - \hat b$,
and that the outer terms can be cast in terms 
of the vector defined above, $\Psi_k =
\frac{1}{N} (\frac{A_k - B_k}{A_k + B_k} )$, yielding
\begin{align}
N \Psi_k - \epsilon_1 < N\hat w_k < N \Psi_k + \epsilon_1,
\end{align}
or equivalently: 
$
\label{eqn:bound_with_phi}
|N \hat w_k - N \Psi_k \ | \le  \epsilon_1$.

This expression 
furnishes us with a coordinate-wise bound for each of the $K$ coordinates of the
geomtric discriminant weight vector, and under the containment event of Lemma 1 we have that they all hold jointly with probability $1-\delta$.
Choosing $\epsilon_1<\epsilon/K$ achieves the requisite bound on the 1-norm of the vector $N \hat w - N\Psi$.

We have established the asymptotic equivalence
between the geometric discriminant weight vector $\hat w$ 
and $\Psi$, defined in terms the solutions to the matrix recurrence in \eqref{eqn:recurstate}.
Next we identify a closed-form 
expression for the diagonalization of $\mathbf{R}$, and thereby for $A_k$,
$B_k$, and $\Psi_k$: with the 
$\vec C_k = \begin{pmatrix}A_k \\ B_k\end{pmatrix}$ and the 
initial conditions $A_0=1$ and $B_0=0$, 
we compute $\vec C_k$ by iterative application of $\mathbf{R}$:
$\vec C_k = \mathbf{R} \vec C_{k-1} = \mathbf{R}^k \vec C_0$. 
To identify a closed form expression for $\mathbf{R}^k$ requires 
a diagonalization of $\mathbf{R}$ such that $\mathbf{R} = {\bf UDU}^{-1}$
for diagonal ${\bf D}$,
and thus $\mathbf{R}^k = {\bf UD}^k{\bf U}^{-1}$.
For ${\bf R} = N \begin{pmatrix}\pin&\pout
\\\pout&\pin \end{pmatrix}$ we
have the following closed form diagonalization, where 
$\mathbf{RU} =\mathbf{DU} $:
\begin{align}
{\bf U} = 
\begin{pmatrix}
-1 & 1
\\
1 &  1
\end{pmatrix}
,
\mathbf{D} = 
\begin{pmatrix}
 \lambda_{1} & 0\\ 
 0 & \lambda_{2}
\end{pmatrix},
\end{align}
where $\lambda_{1}=N(\pin - \pout)$ and $\lambda_{2}=N(\pin + \pout)$.
With the initial  conditions $A_0=1$, $B_0=0$
it follows that the solution to the recurrence relations in \eqref{eqn:recurstate} then simplify to:
\begin{align}
A_{k} =
\dfrac{\lambda_1^k +  \lambda_2^k}{2}
,\hspace{5pt}
B_{k} = 
\dfrac{-\lambda_1^k + \lambda_2^k}{2}
\label{eqn:solrecurequal}.
\end{align}
Finally, we have that $N \Psi_k$ simplifies to 
\begin{align}
N \Psi_k = &
\left(\dfrac{A_k - B_k}{A_{k} + B_{k}}\right) 
=
\dfrac{\lambda_1^k}{\lambda_2^k}
=
\left(
\frac{\pin -\pout}{\pin + \pout}
\right)^k.
\end{align}
Personalized PageRank with $\alpha_* = 
\frac{\pin -\pout}{\pin + \pout}
$ 
employs precisely the weights $(\alpha_*)^k$.
\end{proof}

A few remarks are in order. First, the scalar factor $N$ that differs between the derived weights and $w_{PPR}(\alpha_*)$ does not change the relative ranking of the nodes, since ranking according to the discriminant function $f_1(r^v)=w^Tr^v$ or $f_2(r^v)=Nw^Tr^v$ is equivalent.
Second, the criteria that the stochastic block model be dense ($p_{ij}>0$ and fixed) is necessary for the proof, and it is unclear if a similar result holds for a sparse block model.
Third, the centroid derivations work even when the edge probability 
is higher between blocks than within blocks --- i.e. for disassortative
block models as well as assortative ones.

The simple expression $\alpha_* = (\pin -\pout)/(\pin + \pout)$ 
for the optimal $\alpha$ for geometric classification 
provides a useful interpretation of the choice of $\alpha$
in personalized PageRank: $\alpha$ close to 0 is best for identifying
very strong planted partitions, 
$p_{in} \gg p_{out}$, while $\alpha$ close
to 1 is best when the planted partition is very weak and the
difference $\pin-\pout$ is small.

\vspace{2mm}
\subsection{PageRank and SBMs: $C$ Blocks}
We now state a second proposition that offers a more general connection
between the geometric discriminant weight vector for arbitrary SBMs with
$C$ blocks, the solutions to a matrix recurrence, and 
personalized PageRank. A proof of this more involved
proposition is given in the appendix.
In the $C$-block case, the 
problem is to distinguish between two classes of
nodes: an in-class that may by comprised of nodes in several blocks,
and an out-class comprised of the complement to the in-class. We denote 
these in- and out-classes as $S \subset \{1,\dots, C\}$ and $T = \{1,\dots,C\}
\setminus S$, respectively, and the number of nodes in each class as
$n_{S} = \sum_{i\in S}n_i$ and $n_T= \sum_{i\in T}n_i = n - n_S$.

\vspace{2mm}
\noindent{\bf Proposition 2. }{\it
Let $G_n$ be an $n$-node graph generated from a stochastic block model 
$G(n,\pi,P)$ with $C$ communities and $\pi$ fixed, where
$n_i = \pi_i n$ for $i\in\{1,\dots,C\}$, $\sum_{i=1}^C \pi_i=1$, and $p_{ij}>0$ fixed, $\forall i, j$. 
Let $\hat w$ be the geometric discriminant weight vector in the space
of landing probabilities ($1$-step through $K$-step, $K$ fixed)
between two fixed sets of block classes $S$ and $T$ of $G_n$ of size
$n_S = \sum_{i\in S} n_i$ and
$n_T = \sum_{i\in T} n_i$,
where $S$ and $T$ partition $\{1,\dots,C\}$.

 For any $\epsilon, \delta>0$, there exists an $n$ sufficiently large
such that
$
|| n_{S} \hat w - n_{S} \Psi||_1 \le \epsilon
$
with probability at least $ 1-\delta$, where: 

(a) $\Psi = \Psi(\{x_{i,k}\}_{i=1}^C,n,\pi)$ is a vector with coordinates specified by the
solution to a $C$-dimensional linear homogeneous matrix recurrence relation
$x_{ik} = \sum_{j=1}^C n_i p_{ij} x_{j,k-1}$ for $i\in\{1,\dots,C\}$.

(b) If $\sum_{i\in I} n_i p_{ij} =\sum_{i\in I} n_i p_{ik}$
for each $j,k\in J$
for each $I,J\in \{S,T\}$, then $\Psi$ 
is the solution to a $2$-dimensional linear homogeneous matrix recurrence
and the closed-form solution for $\Psi$ follow from
diagonalizing the corresponding $2\times 2$ matrix.

(c) Assuming that the blocks are identically distributed,
then
$$
n_S \Psi_k 
= \left(\dfrac{(p_{in} -p_{out})}{ C p_{out} + (p_{in} -p_{out})}\right)^k,
$$
in which case the geometric discriminant weight vector
is asymptotically equivalent to personalized PageRank.
}

As with Proposition~\ref{p:1}, this proposition makes heavy use of a concentration lemma 
analogous to Lemma 1, stated and proven as Lemma 2 in the appendix. 
We also employ an additional lemma, Lemma 3, also stated and proven in the appendix.
The result establishes ways in which the optimality of personalized PageRank 
holds even when the two communities have non-trivial substructure.

\begin{figure}[t]
  \centering
\includegraphics[width=0.3\columnwidth]{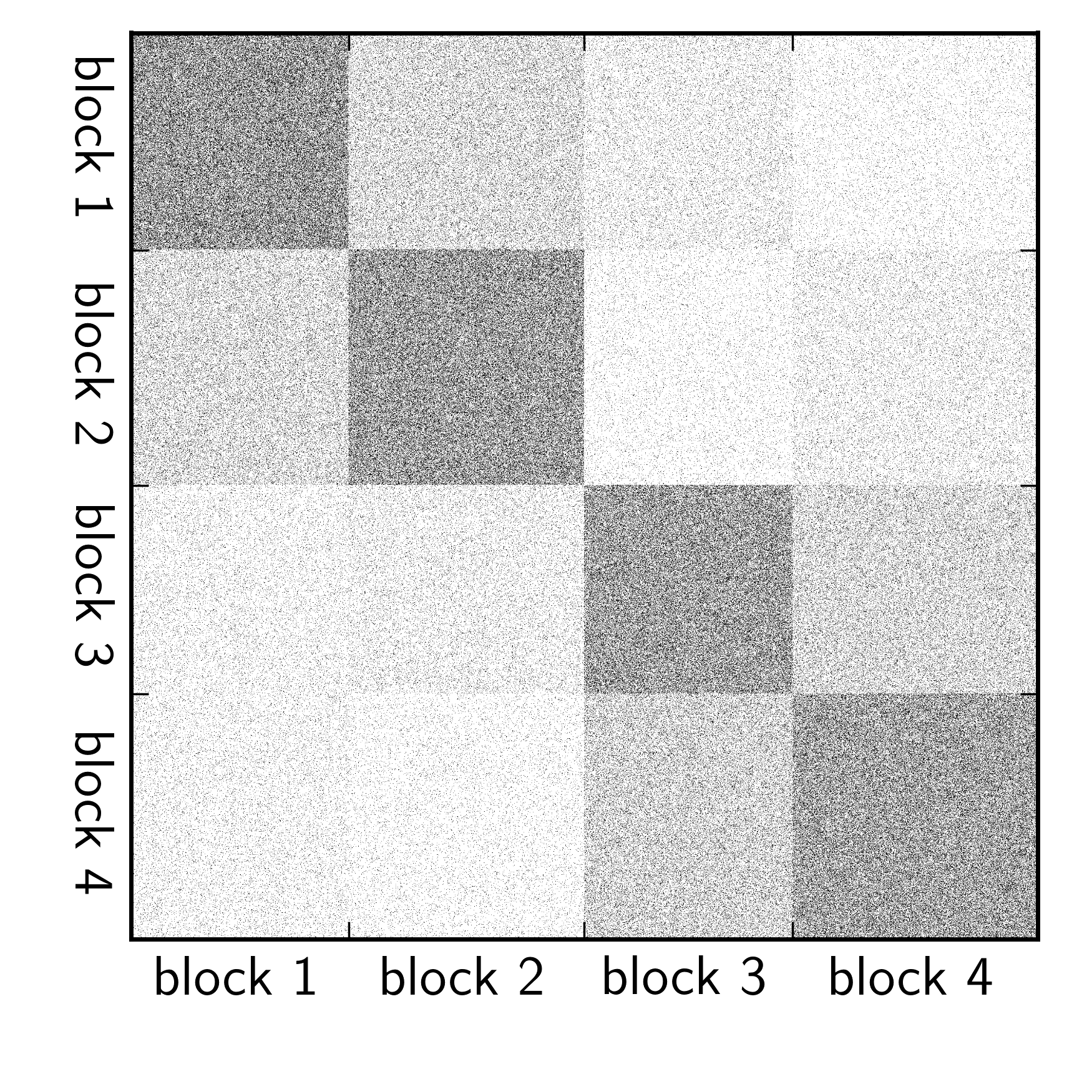}
\hspace{-1pt}
\includegraphics[width=0.3\columnwidth]{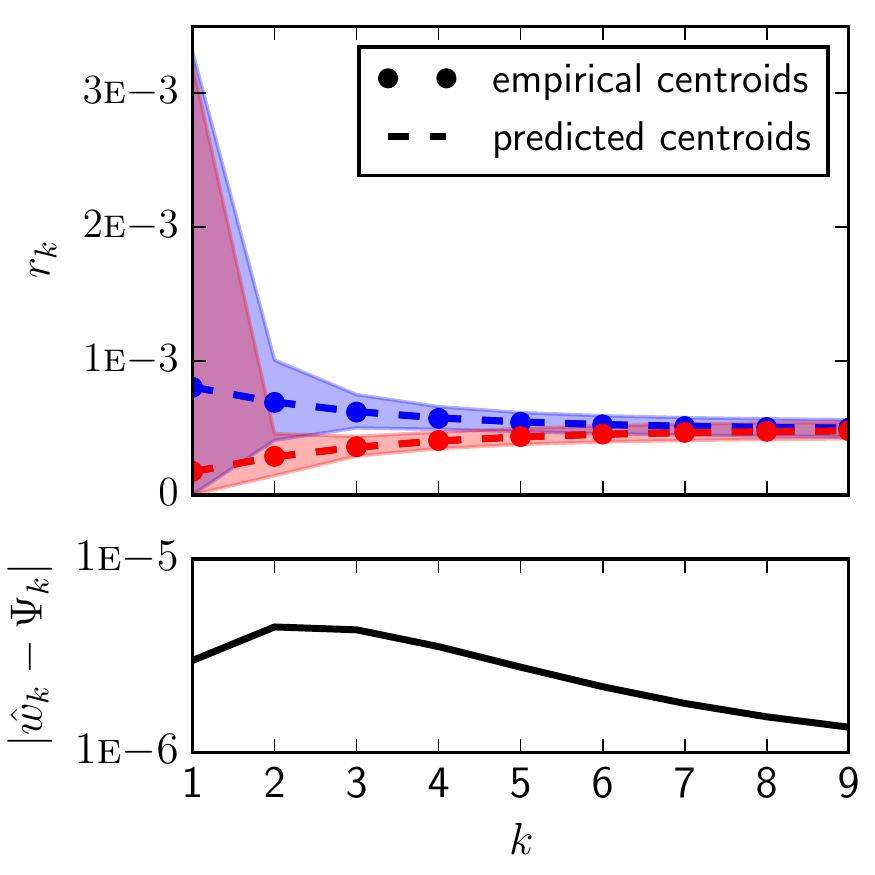}
\vspace{-3mm}
\caption{
Theoretical predictions and empirical calculations for
a 4-block SBM with $n=2048$.
The parameters $\pi_i$ and $p_{ij}$ of the model are given in the appendix
and satisfy Lemma 2 with the block partition $S=\{1,2\}$ and $T=\{3,4\}$.
Left: 
An exemplar adjacency matrix for a single realization of an SBM.
Upper right: theoretical and empirical 
landing probability centroids for the in-class $(a_1,...,a_K)$ (blocks 1 and 2, blue) 
and out-class $(b_1,...,b_K)$ (blocks 3 and 4, red),
shown with empirical [.15\%, 99.85\%] quantiles from 1000 SBM realizations. 
Lower right: difference between the empirically optimal weights $\hat w_k$, computed from the empirical centroids $\hat a$ and $\hat b$ in the upper panel, and our theoretical prediction $\Psi_k$. 
}
\label{fig:sbm_example}
\end{figure}

A major consequence of Proposition 2 is that the asymptotically 
optimal geometric discriminant function
for an arbitrary stochastic block model, without any specification of balanced block
sizes or expected degrees, can be derived from an uncomplicated $C$-dimensional
matrix recurrence relation. Special circumstances lead this recurrence to have a
particularly concise asymptotic form equivalent to personalized PageRank.
But the general solvability of the asymptotically optimal geometric discriminant function
holds for arbitrary block models.

In Figure~1 we see that the asymptotic
centroids show near-perfect agreement with the empirical
centroids for an example block model with $C=4$
non-identical blocks on $n=2048$ nodes ($\pi$ and $P$ are given
 in the appendix).
We also see that the empirical variance of the 
coefficients can be highly
non-uniform, with the 1-step landing probabilities 
exhibiting much greater variance than the
landing probabilities after subsequent steps. This observation
motivates our next approach, where we explicitly consider
these heterogeneous variances, as well as covariances
between the landing probabilities of different step lengths.

\subsection*{Fisherian discriminant functions}
The above geometric approach is a special
case of the more general probabilistic approach to deriving
discriminant functions proposed by Fisher, 
and we will now derive such functions that
consider both the centroids and covariances 
of the sets of landing probabilities.

A Fisherian discriminant function
captures the first two moments (mean and variance) of the 
landing probabilities for each class (the in-class and out-class).
The classes are described by multivariate Gaussians
$N(a, \Sigma_a)$ and $N(b,\Sigma_b)$ 
for the in-class and out-class, respectively.
Here $a$ and $b$ are the same centroids as were derived earlier.
Note that we are not assuming
these point sets obey multivariate Gaussian distributions, 
but we are simply using the Gaussians to capture the first
two moments of the point sets.

Following a standard derivation from quadratic discriminant analysis,
let $z^v \in \{ 0,1\}$ be
the assignment of each node $v \in V$ to one of the two
classes, with $z^v=1$ denoting the seed node class.
The probabilities of a given $r=(r_1,...,r_K)$
belonging to each class are then:
\small
\begin{align}
\Pr(r | z=1) &\propto
|\Sigma_{a}|^{-\frac{1}{2}} \exp \left ( - \frac{1}{2} (r - a)^T \Sigma_{a}^{-1} (r - a) \right ), \\
\Pr(r | z=0 ) &\propto
| \Sigma_{b}|^{-\frac{1}{2}} \exp \left ( - \frac{1}{2} (r - b)^T \Sigma_{b}^{-1} (r - b) \right ). 
\end{align}
\normalsize
Let $\pi_a=\Pr(z=1)$ 
denote the probability that a given
node is in the in-class. When the parameters of the stochastic block
model are known it is clear that $\pi _a= n_a/(n_a+n_b)$.
The log of the likelihood ratio then becomes:
\small
\begin{align}
\nonumber
g(r)&=
\ln \frac{\Pr(r | z=1) \Pr(z=1) }{ \Pr(r | z=0) \Pr(z=0) } \\
\nonumber
&= 
(\underbrace{\Sigma_a^{-1} a - \Sigma_b^{-1} b}_{w})^T r  
+ r^T \underbrace{(\frac{1}{2} \Sigma_{b}^{-1} - \frac{1}{2} \Sigma_{a}^{-1} )}_{W} r 
 + \underbrace{-\frac{1}{2} \left ( a^T \Sigma_a^{-1} a  - b^T \Sigma_b^{-1} b  
+ \ln \frac{| \Sigma_a|}{ | \Sigma_b|}  \right ) + \ln \frac{\pi_a}{1-\pi_a}}_{w_0}
\\
&= w^T r + r^T W r + w_0,
\label{eq:quad}
\end{align}
\normalsize
where we've identified the vector $w$, matrix $W$, and scalar $w_0$ to
simplify notation. In ranking contexts we
we can safely ignore $w_0$, which is constant for all nodes.
\eqref{eq:quad} thus
provides a quadratic discriminant function for ranking in-class 
membership in a manner that accounts for the covariance structure of
the different landing probabilities, e.g. how
the landing probability at a node $u$ after $k$ steps
covaries with the landing probability after $k+1$ steps. 

If we assume $\Sigma_a = \Sigma_b = \sigma^2 I$, we
recover the earlier geometric discriminant function
\begin{align}
\label{eq:g1}
g_{1}(r) = \sigma^{-2} (a - b)^T r + C,
\end{align}
up to an arbitrary additive constant $C$, and observe that the earlier
geometric approach 
corresponds to a uniform and independent variance assumption on
the two point clouds in the space of landing probabilities.
In a slightly more general setting assuming 
$\Sigma_a = \Sigma_b = \Sigma$, meaning that the two 
covariance matrices are identical, 
\eqref{eq:quad} reduces to 
\begin{align}
\label{eq:g2}
g_{2}(r) = [ \Sigma^{-1} (a - b) ]^T r +C,
\end{align}
again up to an arbitrary additive constant $C$.
This discriminant function is still linear, but can have a very different
form than $g_1(r)$. 
While we have shown that personalized PageRank takes
a principled approach to ranking seed community membership, 
the covariance structure of the landing probabilities 
suggests that much better linear discriminant functions 
with the form of \eqref{eq:g2} rather then \eqref{eq:g1} --- 
let alone quadratic functions with the form of \eqref{eq:quad} --- exist
for graphs where the structure can reasonably be 
motivated as coming from a stochastic block model with two classes of blocks.

\subsection*{Learning model parameters}
All the above theoretical derivations assume known parameters, 
but in practical contexts the parameters of the stochastic block model that inform the 
choice of $\alpha$ as well as the covariance matrices must be learned.
Recent results have developed consistent estimators for the parameters
of two-community stochastic block models for an observed graph, 
with two separate estimation regimes: known and unknown block sizes
\cite{frank1982cluster, allman2009identifiability, allman2011parameter,ambroise2012new}.
The former regime admits a closed-form consistent estimator for the edge probabilities,
while the latter regime formulates a tractable composite likelihood function. 
We focus on the former regime of two-block SBMs 
with known block sizes, and will further 
focus on the special case of $p_{aa} = p_{bb} = p_{in}$, $p_{ab} = p_{out}$ 
(also known as the affiliation model) for which known 
consistent estimators $\hat p_{in}$ and $\hat p_{out}$
are reproduced in the appendix.

For the unknown covariance matrices used in Fisherian discriminant functions, 
given the parameter estimates $\hat p_{in}, \hat p_{out}$ and $n_a=n_b$ known, 
we can also estimate the covariance matrices
$\Sigma_a$ and $\Sigma_b$, from simulations of an adequate number of 
stochastic block models with the learned parameters.

\begin{figure*}[t]
\centering
\includegraphics[height=0.153\linewidth]{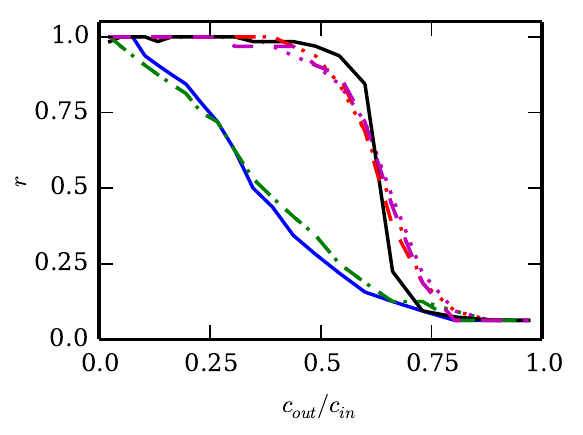}
\includegraphics[height=0.153\linewidth]{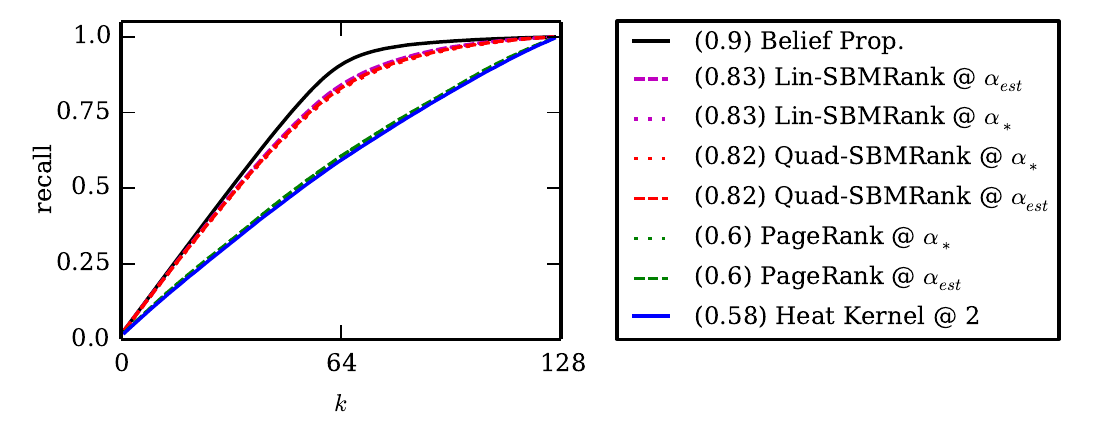}
\includegraphics[height=0.153\linewidth]{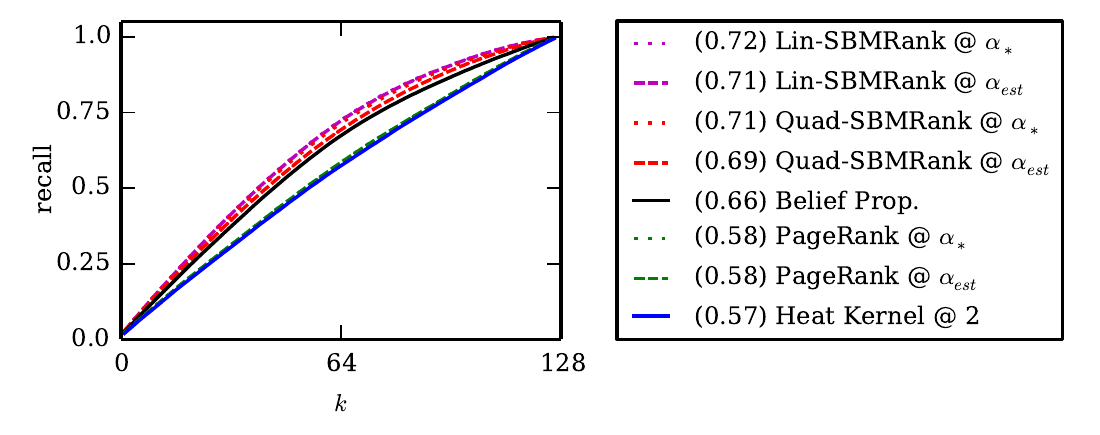}
\vspace{-3mm}
\caption{
The performance of various seed set expansion classifiers: Belief Propagation (black),
personalized PageRank (green), and heat kernel (blue), Lin-SBMRank (magenta), and Quad-SBMRank (red).
Left:
The Pearson correlation $r$ between the recovered
labeling and the true labeling for a SBM on $n=128$ nodes, with $n_a=n_b=64$ 
and expected degree $\bra d \ket = 16$,
as a function of the
out/in balance $c_{out}/c_{in}$.
Right:
The cumulative recall of two stochastic block
models corresponding to $c_{out}/c_{in} = 0.6$ and $0.66$, respectively, 
from the left hand panel
(or $p_{in}=0.3$,
$p_{out}=0.2$ and
$p_{in}=0.3125$, $p_{out}=0.1875$).
The legends indicate the recalls at $k=64$.
}
\label{f:sbm_results}
\end{figure*}

\section{Computational Results} 
We now evaluate our optimal geometric and Fisherian approaches to 
seed set expansion on graphs that have 
been generated by stochastic block models, where 
ground truth community structure is defined
from the generative process.

We begin by examining
how well classifications based on our
discriminant functions correlate with the underlying
true partitioning of the SBMs being considered,
a standard test of inference methods for SBMs.
While exact recovery of planted partitions has been
shown possible in regimes where the the two blocks
are well separated (where $|p_{in}-p_{out}|$ is
large enough) \cite{condon2001algorithms,dyer1989solution,snijders1997estimation},
recent work has shown that recovering a partition
that is correlated with the underlying partition is
still possible even in some contexts when exact recovery
is impossible, up to a recently identified resolution limit 
\cite{decelle2011asymptotic,mossel2014belief}.

Figure~\ref{f:sbm_results}A shows the
Pearson correlation $r$ between the recovered
partition and the ground truth partition
for various discriminant functions on a SBM 
with $n=128$ nodes, $n_a=n_b$,
and average degree $\bra d \ket =16$.
Our covariance-adjusted methods recover
a correlated partition up to the same resolution limit as Belief Propagation, 
and they slightly out-perform Belief Propagation in the
regime just beyond the resolution limit.
Meanwhile personalized PageRank (without
any covariance adjustment)
and heat kernel perform poorly by comparison.

In Figure~\ref{f:sbm_results}B and C we see 
cumulative recall curves for a SBM 
with $n_a=n_b=64$ nodes and two choices of $p_{in}$ and $p_{out}$ 
(given in the caption).
The personalized PageRank classification 
is ranked according 
to \eqref{eq:g1}, whereas 
Lin-SBMRank and Quad-SBMRank have the forms of
\eqref{eq:g2} and \eqref{eq:quad}
respectively.
The recall curves measure the recall of 
the classification methods seeded with a single node seed set and
attempting to return a seed block of $m$ nodes, 
as a function of $m$.
We see that that our quadratic discriminant function
has considerably improved recall over
ordinary personalized PageRank,
identifying the first 64 nodes with high recall. 
In contrast we see that personalized PageRank 
(with either $\alpha_{est}$, based on estimated parameters,
 or $\alpha_*$, based on the true parameters) and heat kernel
(with $t=2$) perform the task with comparable recalls. 
Our linear
method that utilizes a single covariance matrix
for the two classes exhibits a recall nearly
identical to the fully quadratic method.

Our interest here is on
classifications in the space of
landing probabilities. 
Belief Propagation, which does not work within this 
framework,
has been recently shown to achieve correlated
classification up to the resolution limit 
\cite{decelle2011asymptotic,mossel2014belief,zhang2014scalable},
as have modified spectral methods
applied to different non-standard analogs of the Laplacian
matrix known as the non backtracking matrix \cite{krzakala2013spectral} 
and the Bethe-Hessian \cite{saade2014spectral}. 
We thus take Belief Propagation
as a benchmark for optimal performance, but at
the cost of additional computation complexity.
For a more comprehensive discussion of 
Belief Propagation, see the appendix. 
The random walks that underly our
landing probabilities are standard uniform random walks,
and it is an interesting 
open question whether
classifiers in the space
of landing probabilities of alternative random walks 
(e.g.~non-backtracking random walks) may yield even better approaches. 

\section{Discussion}
This work contributes a principled motivation
for personalized PageRank's success
in contextually ranking nodes in graphs.
Personalized PageRank and heat kernel methods
both rank using linear
discriminant functions in the space of random walk landing probabilities.
We show that personalized PageRank in fact arises as the optimal
geometric discriminant function in the space of 
landing probabilities for classifying 
nodes in a hidden seed community
in a stochastic block model.
Building on this connection between stochastic block
models and personalized PageRank, we develop
more complex covariance-adjusted linear and quadratic 
approaches to classification in the space
of landing probabilities.
We see that these classifiers dramatically outperforms personalized PageRank
and heat kernel methods for recovering seed sets in 
graphs generated from stochastic block models. 

The connection between personalized PageRank
and stochastic block models 
is surprising, and we see it pointing toward
a wide range of related research questions. 
Can the recent rigorous results for the resolution
limit of stochastic block models \cite{mossel2014belief} 
provide insights into a broader
class of contextual ranking problems? 
Is there an alternative classification algorithm (or alternative graph model)  
where heat kernel methods emerge as optimal?
Can other new or existing machine learning or ranking methods be motivated
through principled analyses of structured models?
There are also a host of further questions that would serve
to improve the details of the specific approach we outline here.
Can the joint distribution of random walk landing probabilities be 
modeled more explicitly than by a multivariate Gaussian 
(approximating just the first two moments)? 
The potential application of our quadratic discriminant classifier
to diverse contextual ranking problems
suggests revisiting the broad range of applied problems where
PageRank has found success.

\appendix

\newcommand{\beginsupplement}{%
        \setcounter{table}{0}
        \renewcommand{\thetable}{S\arabic{table}}%
        \setcounter{figure}{0}
        \renewcommand{\thefigure}{S\arabic{figure}}%
     }
\beginsupplement

\section{Additional Proofs for 2--dimensional Case}
\subsection{Proof of Lemma 1}
\vspace{2mm}
\noindent{\bf Lemma 1. }{\it
Let $G_n$ be an $n$-node graph generated from a two-block stochastic block model 
$G(n,\pi,P)$ with equally sized communities ($\pi_1=\pi_2=1/2$), $N= n/2$, and
$P$ fixed with $p_{ij}>0$, $\forall i, j$, where the first block is designated the seed block.

For any $\epsilon, \delta>0$,
there is an $n$ sufficiently large
such that the random landing probabilities 
$(\hat a_1,....,\hat a_K)$ 
and $(\hat b_1,...,\hat b_K)$
for a uniform random walk on $G_n$ starting in the seed block
satisfy the following conditions with probability at least $1-\delta$ for 
all $k>0$:
\begin{align}
N \hat a_k 
& \in 
  \left[(1 - \epsilon) 
  \frac{A_k}{A_k + B_k}
,
  (1 + \epsilon) 
  \frac{A_k}{A_k + B_k}
\right]
\text{ and  }
\\
N \hat b_k
& \in  
  \left[(1 - \epsilon) 
  \frac{B_k}{A_k + B_k}
  ,
  (1 + \epsilon) 
  \frac{B_k}{A_k + B_k}
  \right],
\end{align}
where $A_k$, $B_k$ are the solutions to the
matrix recurrence relation
\begin{eqnarray}
  \left\{
\begin{aligned}
A_k &= N (\pin A_{k-1} + \pout A_{k-1})\\
B_k &= N (\pout B_{k-1} + \pin B_{k-1}),
\end{aligned}
\right.
\label{eqn:recurstateA}
\end{eqnarray}
with $A_0=1$, $B_0=0$.
}
\begin{proof}

We first introduce some useful notation. Let $V_a$ denote the set of nodes in the seed block and $V_b$ denote the complement.
We define the following walk counts from the seed
to each node, which are random variables under the
randomness of the block model:
\begin{align*}
\inpathsVar^u_k =  \text{\# paths from $s$ to $u \in V_a$ of length $k$},\\
\outpathsVar^u_k = \text{\# paths from $s$ to $u \in V_b$ of length $k$}.
\end{align*}
The seed node $s\in V_a$ is given and fixed, and therefore suppressed in our notation. 
We denote the number of walks of length $k$ originating at $s$ and ending in $V_a$ 
and $V_b$, respectively, as:
\begin{align}
\textstyle 
\inpathsVar_k = \sum_{u \in V_a} \inpathsVar^u_k, \hspace{1cm}
\outpathsVar_k = \sum_{u \in V_b} \outpathsVar^u_k.
\end{align}
We see then that the random aggregate landing probabilities,
the probabilities that a $k$-step walk starting at a seed node
in $V_a$ ends in $V_a$, and the probability that it ends in $V_b$, are then:
\begin{align}
\inprobrandVar_k 
= \frac{1}{n_a}\frac{\inpathsVar_k}{\inpathsVar_k + \outpathsVar_k},
\hspace{0.5cm}
\outprobrandVar_k 
= \frac{1}{n_b} \frac{\outpathsVar_k}{\inpathsVar_k + \outpathsVar_k}.
\end{align}

Our proof strategy is to show that these quantities $\hat a_k$ and $\hat b_k$ 
concentrate around expressions
given in terms of the solutions $A_k$ and $B_k$ of a recurrence. 
We then show concentration for the quantities
$A_k/(A_k+B_k)$ and $A_k/(A_k+B_k)$; notably, they concentrate
around values that are not their expectations.

An obstruction to simply taking the expectations
of the walk counts $E[\hat a_k]$ and $E[\hat b_k]$
(and showing
concentration around the ratio of expectations)
is that counting length-$k$ walks for $k>1$ requires 
counting walks that possibly revisit edges, creating
a dependence between walk counts of different lengths.
The recurrence solutions $A_k$ and $B_k$
that we will analyze can in fact be thought
of as the expected walk counts on a slightly different random
graph model,
where the edges are independently resampled after each walk step.
What our analysis effectively shows is that the
walk counts on the stochastic block model (our model of interest)
concentrate on the expected walk counts of that alternative model
(where edges are independently resampled at after each walk step).
This connection between models is mentioned only as an optional 
pedagogical tool, and is not essential to understanding our proof.

We now return to the walk-count random variables $\hat a_k$ and $\hat b_k$
in a graph $G_n$ drawn from the stochastic block model.
Suppose we are given $\epsilon > 0$ and $\delta > 0$
as in the statement of the lemma, and we seek bounds
for a specific walk length $k\le K$.
We choose $\gamma_2 > 0$ small enough that
$(1 - \gamma_2) / (1 + \gamma_2) \geq 1 - \epsilon$ and 
$(1 + \gamma_2) / (1 - \gamma_2) \leq 1 + \epsilon$;
we then choose $\gamma$ small enough that
$(1 - \gamma)^K \geq 1 - \gamma_2$
and $(1 + \gamma)^K \leq 1 + \gamma_2$.

Let $\hat M_{uv}$ be a matrix of independent Bernoulli random variables,
indicating the edge event when $(u,v)$ is an edge in the graph $G_n$. 
Notice that $\sum_{u\in V} \hat M_{uv}$ is the random 
out-degree of node $v$.
We observe that each node $v\in V_a$ has
in expectation a total of
$d_{aa}$ or $d_{ab}$ edges to nodes in $V_a$ or $V_b$, where
\begin{align}
\label{eqn:expecteddegs}
d_{aa} 
= \mathbb E \left[\sum_{u\in V_a} \hat M_{uv}\right] 
= N p_{in}, \hspace{3mm}
d_{ab} 
= \mathbb E \left[\sum_{u\in V_a} \hat M_{uv}\right] 
= N p_{out},
\end{align}
and similarly $d_{ba} = N p_{out}$ and $d_{bb} = N p_{in}$.
When the expectations $p_{in},p_{out}$ are fixed in $n$ we can use standard multiplicative Chernoff bounds to bound
the probabilities of $4n$ bad events. We have that for any $\gamma>0$ and any $i$, $j \in \{a,b\}$:
\begin{align}
\Pr \left ( \sum_{u\in V_i} \hat M_{uv} \notin [(1-\gamma)d_{ij} , (1+\gamma) d_{ij}] \right ) \le C_1 e^{-n} 
\end{align}
for some constant $C_1$ for any $v \in V_j$.
Across all $i$, $j$ pairs there are $4n$ bad events, and we want to lower bound the probability of there being no bad event. By the union bound we have that
\begin{align}
\Pr \left ( \sum_{u\in V_i} \hat M_{uv} \in [(1-\gamma)d_{ij} , (1+\gamma) d_{ij}], \forall v \in V_j,  \forall i, j \right ) \ge 1 - 4C_1n e^{-n} .
\end{align}
Thus, it is clear that for any $\gamma>0$ and any $\delta>0$, there exists an $n$ sufficiently large such that 
the probability that none of the degrees exceed a multiplicative factor of $(1\pm \gamma)$ is at least $1-\delta$.
Assuming this containment succeeds, which is to say assuming
\begin{align}
\label{eq:contains2}
\sum_{u\in V_i} \hat M_{uv} \in [(1-\gamma)d_{ij} , (1+\gamma) d_{ij}], 
\hspace{3mm} 
\forall v \in V_j, \forall i, j,
\end{align}
then the rest of the proof argument is deterministic.
The next step of our proof strategy is to show that we also have
\begin{align}
\label{eq:AandB}
\hat A_k \in [(1-\gamma_2) A_k , (1+\gamma_2)A_k]\ \text{ and } \ 
\hat B_k \in [(1-\gamma_2)B_k , (1+\gamma_2)B_k]\ 
\end{align}
whenever the stated containment event holds.

We give a proof by induction. 
First we define a new set of variables:
\begin{align}
\hat H_k^u = 
\begin{cases}
\hat A_k^u \text{ if } u \in V_a,\\
\hat B_k^u \text{ if } u \in V_b.
\end{cases}
\end{align}
We then begin with the base case, furnishing an upper bound on $\hat A_1$:
\begin{align}
\hat A_1 =& \sum_{u\in V_a} \hat H_1^u 
= \sum_{u \in V_a}\sum_{v\in V} \hat M_{u v} \hat H^v_{0} \\
=& \sum_{v\in V_a}\hat H^v_{0} \sum_{u \in V_a} \hat M_{u v} 
+ \sum_{v\in V_b}\hat H^v_{0} \sum_{u \in V_a} \hat M_{u v} \\
\le&
\sum_{v\in V_a}\hat H^v_{0} (1+\gamma) d_{aa}
+ \sum_{v\in V_b}\hat H^v_{0} (1+\gamma)d_{ab} \\
=& (1+\gamma)d_{aa} = (1+\gamma) A_1. 
\end{align}
Notice that in the last step we have identified 
the expectation in \eqref{eqn:expecteddegs}
and 
$A_1$ from \eqref{eqn:recurstateA} in the lemma statement.
Using a similar set of steps one can easily see that $(1-\gamma)A_1 \le \hat A_1$ 
and $(1-\gamma) B_1 \le \hat B_1 \le (1+\gamma) B_1$ also
hold.

Next, for our induction we assume that 
\begin{align}
\hat A_k \in [(1-\gamma)^k A_k, (1+\gamma)^k A_k],\\
\hat B_k \in [(1-\gamma)^k B_k, (1+\gamma)^k B_k],
\end{align}
and want to show that the above implies that
\begin{align}
\hat A_{k+1} \in [(1-\gamma)^{k+1} A_{k+1}, (1+\gamma)^{k+1} A_{k+1}],\\
\hat B_{k+1} \in [(1-\gamma)^{k+1} B_{k+1}, (1+\gamma)^{k+1} B_{k+1}].
\end{align}
We upper-bound $\hat A_{k+1}$:
\begin{align}
\hat{A}_{k+1} =&  \sum_{u \in V_a}\hat H_{k+1}^u 
= \sum_{u \in V_a}\sum_{v\in V} \hat M_{u v} \hat H^v_{k} \\
=& \sum_{v\in V_a}\hat H^v_{k} \sum_{u \in V_a} \hat M_{u v} 
+ \sum_{v\in V_b}\hat H^v_{k} \sum_{u \in V_a} \hat M_{u v}\\
\le &\ \hat A_k (1+\gamma)d_{aa} + \hat B_k (1+\gamma) d_{ab}\\
\le &\ (1+\gamma)^{k+1} A_{k} d_{aa} + (1+\gamma)^{k+1} B_{k} d_{ab} \\
=& (1+\gamma)^{k+1} A_{k+1}, 
\end{align}
where in the last inequality we use the induction hypothesis 
and
the last equality again identifies \eqref{eqn:recurstateA} from the lemma statement.
We observe that $\hat A_{k+1}\le (1+\gamma)^{k+1} A_{k+1}$, and
similar steps furnish the lower bound $ (1-\gamma)^{k+1} A_{k+1} \le \hat A_{k+1}$
and that $ (1-\gamma)^{k-1} B_{k+1} \le \hat B_{k+1} \le
 (1-\gamma)^{k+1} B_{k+1}$,  
completing the proof by induction. 

As a result, we have:
\begin{align}
\hat A_k \in [(1 - \gamma)^kA_{k} , (1+\gamma)^kA_{k}] \text{ and }\
\hat B_k \in [(1 - \gamma)^kB_{k} , (1+\gamma)^kB_{k}].
\end{align}
Since $\gamma$
was chosen such that $(1 - \gamma)^K \geq 1 - \gamma_2$
and $(1 + \gamma)^K \leq 1 + \gamma_2$ we then have that
\begin{align}
\hat A_k \in [(1 - \gamma_2)A_{k} , (1+\gamma_2)A_{k}]
\text{ and }
\hat B_k \in [(1 - \gamma_2)B_{k} , (1+\gamma_2)B_{k}],
\end{align}
as desired in \eqref{eq:AandB}.
We then also have that for $\hat A_k + \hat B_k$,
\begin{align}
\hat A_k + \hat B_k \in&\ [(1 - \gamma_2)(A_{k} +B_{k}), 
(1+\gamma_2)(A_{k}+B_{k})].
\end{align}
Finally, since $\epsilon$ satisfies
$(1 - \gamma_2) / (1 + \gamma_2) \geq 1 - \epsilon$ and 
$(1 + \gamma_2) / (1 - \gamma_2) \leq 1 + \epsilon$ 
based on our choice of $\gamma_2$
we have
\begin{align}
\frac{\hat{A}_k}{\hat{A}_k + \hat{B}_k} \in 
  [(1 - \epsilon) 
  \frac{A_{k}}{A_{k} +B_{k}}
  , 
  (1 + \epsilon) 
  \frac{A_{k}}{A_{k} + B_{k}}],
\\
\frac{\hat{B}_k}{\hat{A}_k + \hat{B}_k} \in 
  [(1 - \epsilon) 
  \frac{B_{k}}{A_{k} +B_{k}}
  , 
  (1 + \epsilon) 
  \frac{B_{k}}{A_{k} +B_{k}}].
\end{align}
These final containments hold whenever the original containment in \eqref{eq:contains2} holds,
with probability at least $1-\delta$, completing the proof.

\end{proof}

Notice that a closed-form expression for 
$\vec C_k = \begin{pmatrix}A_k \\ B_k\end{pmatrix}$ 
can be obtained by
diagonalizing $\mathbf{R}=\begin{pmatrix}
N \pin & N\pout \\
N\pout & N\pin
\end{pmatrix}$,
that is, identifying $\mathbf{U},\mathbf{D},\mathbf{U}^{-1}$ for diagonal $\mathbf{D}$ such that
$\mathbf{R} =\mathbf{ UDU}^{-1}$ and thus $\vec C_k = U D^k U^{-1} \vec C_0$.

\section{Proofs for C--dimensional Case}

In this section of the appendix we build up a 
more general connection between stochastic block models
on $C>2$ blocks, 
geometric classification in the space of landing probabilities, and personalized PageRank.
The main result of the section is Proposition 2, which relies on two lemmas. The first lemma,
Lemma 2, is a concentration lemma analogous to Lemma 1. The second lemma,
Lemma 3, provides useful sufficient conditions for a closed form diagonalization employed
in the proof of the ultimate proposition.

To further motivate Lemma 3 we note that it is non-trivial to 
identify a closed form diagonalization of the general $C\times C$ matrix $R$.
For the geometric discriminant weight vector we can 
identify the two aggregate quantities $\sum_{i\in S}x_{ik}$
 and $\sum_{i\in T}x_{ik}$, which correspond to the aggregate 
number of walks landing in $S$ versus $T$; in particular
 we do not need, as ends themselves, the walk counts to each individual block
$\{x_{ik}\}_{i=1}^C$. 
Under the specified degree homogeneity conditions (on the parameters 
characterizing blocks in the two sets, $S$ and $T$),
we can directly
formulate a $2\times2$ recurrence relation
for
$\sum_{i\in S}x_{ik}$ and
$\sum_{i\in T}x_{ik}$, and circumvent the $C\times C$ diagonalization.
Lemma 3 makes this correspondence precise
by demonstrating that
when all nodes have the same expected degree and number of edges
to nodes in the in-class, 
we may pivot from studying a $C$-dimensional recurrence for
$\{x_{ik}\}_{i=1}^C$ to a $2$-dimensional recurrence in terms of the 
aggregate quantities $\sum_{i\in S}x_{ik}$ and
$\sum_{i\in T}x_{ik}$. 

We leverage the simplification sanctioned by Lemma 3 to
identify a closed form solution for $\Psi$
and observe that under the stated additional conditions, 
the geometric discriminant weight vector for the $C$-block
problem is again asymptotically equivalent to personalized PageRank.

\subsection{Proof of Lemma 2}
\vspace{2mm}
\noindent{\bf Lemma 2. }{\it
Let $G_n$ be an $n$-node graph generated from a stochastic block model 
$G(n,\pi,P)$ with $C$ communities and $\pi$ fixed, where
$n_i = \pi_i n$ for $i\in\{1,\dots,C\}$, $\sum_{i=1}^C \pi_i=1$, and $p_{ij}>0$ fixed, $\forall i, j$. 

For any $\epsilon, \delta>0$,
there is an $n$ sufficiently large
such that the random landing probabilities 
$(\hat y_i^1,....,\hat y_i^K)$ for $i\in{1,\dots,C}$
for a uniform random walk starting at a uniformly random node in block $j$
with probability $x_{j0}$
satisfy the following conditions with probability at least $1-\delta$ for 
all $k>0$:
\begin{align}
n_i \hat y_{ik} 
\in &
  \left[(1 - \epsilon) 
  \frac{x_{ik}}{\sum_{j=1}^C x_{jk}}
,
  (1 + \epsilon) 
  \frac{x_{ik}}{\sum_{j=1}^C x_{jk}}
\right]
\text{ for } i\in\{1,\dots,C\},
\label{eqn:containstate}
\end{align}
where $\left\{x_{ik}\right\}_{i=1}^C$ are the solutions to the
matrix recurrence relation
\begin{align}
x_{ik} = \sum_{j = 1}^C R_{ij} x_{j,k-1} \text{ for } i\in\{1,\dots, C\}.
\label{eqn:recurstate2}
\end{align}
with initial conditions $\left\{x_{j0} \right \}_{j=1}^C$. 
}

\begin{proof}
We again begin by introducing some useful notation.
Let the node set $V$ be composed of $C$ disjoint subsets $V_1,...,V_C$ constituting the nodes in the $C$ blocks of the model.
We define the following walk counts from a seed node $s$
to each node, which are random variables under the
randomness of the block model:
\begin{align*}
\hat X^u_{ik} =  \text{\# paths from $s$ to $u \in V_i$ of length $k$}, \hspace{2mm} \text{ for }i\in \{1,\dots, C\}.
\end{align*}
The seed node $s\in V_j$ for some 
distinguished 
$j \in \{1,\dots,C\}$ is given and fixed, and therefore suppressed in our notation. 
We denote the number of walks of length $k$ originating at $s$ and ending in $V_i$ for $i\in\{1,\dots,C\}$ as:
\begin{align}
\textstyle 
\hat X_{ik} = \sum_{u \in V_i} \hat X^u_{ik}.
\end{align}
We see that the random aggregate landing probabilities,
the probabilities that a $k$-step walk starting at a seed node
in $j$ ends in community $i$, are then:
\begin{align}
\hat y_{ik}
= \frac{1}{n_i}\frac{\hat X_{ik}}{\sum_{j=1}^C \hat X_{jk}}.
\end{align}

Our proof strategy is to show that these quantities, $\{\hat y_{ik}\}_{i=1}^C$, concentrate around expressions
given in terms of the solutions $\{x_{ik}\}_{i=1}^C$.
The values $\left\{x_{ik}/\sum_{j=1}^C x_{jk}\right\}_{i=1}^C$ 
that we will show the $\left\{\hat y_{ik}\right\}_{i=1}^C$ concentrate around are notably not their expectations. 

As in the two-block case, an obstruction to simply taking the expectations
of the walk counts $\left\{\hat X_{ik}\right\}_{i=1}^C$ (and showing
concentration around the ratio of expectations)
is that counting length-$k$ walks for $k>1$ requires 
counting walks that possibly revisit edges, creating
a dependence between walk counts of different lengths.
The recurrence solutions $\{x_{ik}\}_{i=1}^C$ 
that we will analyze can in fact be thought
of as the expected walk counts on a slightly different random
graph model,
where the edges are independently resampled after each walk step.
What our analysis effectively shows is that the
walk counts on the stochastic block model, our model of interest,
concentrate on the expected walk counts of that alternative model.
As in the proof of Lemma 1,
this connection between models is mentioned only as an optional 
pedagogical tool, and is not essential to understanding our proof.

In Lemmas 3 we introduce the recurrence relations:
\begin{align}
x_{ik} =& \sum_{j = 1}^C R_{ij} x_{j,k-1} \text{ for } i\in\{1,\dots, C\},\\
&\text{where }x_{i0} = \begin{cases}1 &\tif i=i_S \\ 0 &\telse\end{cases}
\end{align}
and establish 
that under certain conditions on the parameters of the SBM
we can identify general closed-form solutions for $\bf{U}$, $\bf{D}$, and $\bf{U}^{-1}$ where
$\bf{UDU}^{-1} = \bf{R}$, which allows us to
obtain a closed form solution for $\left\{x_{ik}\right\}_{i=1}^C$.
We emphasize, however, that the containments in \eqref{eqn:containstate} proved
herein are valid even when the closed-form expressions for $\{x_{ik}\}_{i=1}^C$ are not known, i.e.~even when a closed-form diagonalization of $\bf {R}$ has not been
identified.

We now return to the walk count random variables $\left\{\hat X_{ik}\right\}_{i=1}^C$ 
in a graph $G_n$ drawn from the stochastic block model.
Suppose we are given $\epsilon > 0$ and $\delta > 0$
as in the statement of the lemma, and we seek bounds
for a specific walk length $k\le K$.
We choose $\gamma_2 > 0$ small enough that
$(1 - \gamma_2) / (1 + \gamma_2) \geq 1 - \epsilon$ and 
$(1 + \gamma_2) / (1 - \gamma_2) \leq 1 + \epsilon$;
we then choose $\gamma$ small enough that
$(1 - \gamma)^K \geq 1 - \gamma_2$
and $(1 + \gamma)^K \leq 1 + \gamma_2$.

Let $\hat M_{uv}$ be a matrix of independent Bernoulli random variables,
indicating the edge event when $(u,v)$ is an edge in the graph $G_n$. 
Notice that $\sum_{u\in V} \hat M_{uv}$ is the random 
out-degree of node $v$.
We observe that for each $j \in \{1,\dots,C\}$, each node $v\in V_j$ has
in expectation a total of
$d_{ij}$ edges to nodes in $V_i$, where
$$d_{ij}  
= \mathbb E \left[\sum_{u\in V_i} \hat M_{uv}\right] 
= \sum_{u\in V_i} p_{ij} 
= n_i p_{ij}.
$$
When the expectations $p_{ij}$ are fixed in $n$ we can use standard multiplicative Chernoff bounds to bound
the probabilities of $4n$ bad events. We have that for any $\gamma>0$ and any $i$, $j \in \{1,\dots,C\}$:

\begin{align}
\label{eq:Mbound}
\Pr \left ( \sum_{u\in V_i} \hat M_{uv} \notin [(1-\gamma)d_{ij} , (1+\gamma) d_{ij}] \right ) \le C_1 e^{-n} 
\end{align}
for some constant $C_1$ for any $v \in V_j$.
Across all $i$, $j$ pairs there are $4n$ bad events, and we want to lower bound the probability of there being no bad event. By the union bound we have that
\begin{align}
\Pr \left ( \sum_{u\in V_i} \hat M_{uv} \in [(1-\gamma)d_{ij} , (1+\gamma) d_{ij}], \forall v \in V_j,  \forall i, j \right ) \ge 1 - 4C_1n e^{-n} .
\end{align}
Thus, it is clear that for any $\gamma>0$ and any $\delta>0$, there exists an $n$ sufficiently large such that 
the probability that none of the degrees exceed a multiplicative factor of $(1\pm \gamma)$ is at least $1-\delta$.
Assuming that this containment succeeds, which is to say assuming 
\begin{align}
\label{eq:contains}
\sum_{u\in V_i} \hat M_{uv} \in [(1-\gamma)d_{ij} , (1+\gamma) d_{ij}], \forall v \in V_j, \forall i, j,
\end{align}
then the rest of the proof is deterministic. 
The next step of our proof strategy is to show that we also have
\begin{align}
\label{eq:AandB2}
\hat X_{ik} \in [(1-\gamma_2) x_{ik} , (1+\gamma_2)x_{ik}]\ \text{ for } i\in\{1,\dots,C\} \ 
\end{align}
whenever the stated containment event holds.

We give a proof by induction. 
First we define a new set of variables:
\begin{align}
\hat H_k^u = 
\begin{cases}
\hat X_{ik}^u \text{ if } u \in V_i,\\
\end{cases}
\text{ for } i\in \{1,\dots,C\}.
\end{align}
We then begin with the base case, furnishing an upper bound on $\hat X_{i1}$ for each $i\in\{1,\dots ,C\}$:
\begin{align}
\hat X_{i1} =& \sum_{u\in V_i} \hat H_1^u 
= \sum_{u \in V_i}\sum_{v\in V} \hat M_{u v} \hat H^v_{0} 
= \sum_{j=1}^C \sum_{v\in V_j} \hat X^v_{j0}  \sum_{u\in V_i} \hat M_{u v} \\
\le &
(1+\gamma)\sum_{j=1}^C \sum_{v\in V_j} \hat X^v_{j0}\ d_{ij}
= (1+\gamma)\sum_{j=1}^C d_{ij} x_{j0}
= (1+\gamma) x_{i1}
\end{align}
where in the final steps we used the initial conditions that imply 
$\sum_{v\in V_j} \hat X_{j0}^v= x_{j0}$.
Using a similar set of steps one can easily see that $\hat
X_{i1}\ge(1-\gamma)x_{i1} $ for each $i\in\{1,\dots,C\}$ also holds.

Next, for our induction we assume that 
\begin{align}
\hat X_{ik} \in [(1-\gamma)^k x_{ik}, (1+\gamma)^k x_{ik}] \text{ for } i\in\{1,\dots,C\},
\end{align}
and want to show that the above implies that
\begin{align}
\hat X_{i,k+1} \in [(1-\gamma)^{k+1} x_{i,k+1}, (1+\gamma)^{k+1} x_{i,k+1}] \text{ for } i\in\{1,\dots,C\}.
\end{align}
We upper-bound $\hat X_{i,k+1}$:
\begin{align}
\hat{X}_{i,k+1} =&  \sum_{u \in V_i}\hat H_{k+1}^u 
=\sum_{j=1}^C \sum_{v\in V_j}\hat X^v_{jk} \sum_{u \in V_i} \hat M_{u v} \\
\le&(1 + \gamma) \sum_{j=1}^C \sum_{v\in V_j}\hat X^v_{jk} d_{ij} 
=(1+ \gamma) \sum_{j=1}^C \hat X_{jk} d_{ij} \label{eqn:usechernoff}\\
\le&(1+ \gamma)^{k+1} \sum_{j=1}^C x_{jk} d_{ij} 
= (1+\gamma)^{k+1} x_{i,k+1},\label{eqn:useinduc}
\end{align}
where for the upper bound in \eqref{eqn:usechernoff} we use the containment guarantee 
from \eqref{eq:contains}, and for the upper bound in \eqref{eqn:useinduc}
we use the induction hypothesis. 
We observe that $\hat X_{i,k+1}\le (1+\gamma)^{k+1} x_{i,k+1}$, and
similar steps furnish the lower bound $\hat X_{i,k+1}\ge (1-\gamma)^{k+1} x_{i,k+1} $ for each 
$i\in\{1,\dots,C\}$,
completing the proof by induction. 

As a result, we have:
\begin{align}
\hat X_{ik} \in \left[(1 - \gamma)^k x_{ik} , (1+\gamma)^kx_{ik}\right] \text{ for } i\in\{1,\dots,C\}.
\end{align}
Since $\gamma$ 
was chosen such that $(1 - \gamma)^K \geq 1 - \gamma_2$
and $(1 + \gamma)^K \leq 1 + \gamma_2$ we then have that
\begin{align}
\hat X_{ik} \in \left[(1 - \gamma_2)x_{ik} , (1+\gamma_2)x_{ik}\right] \text{ for } i\in\{1,\dots,C\},
\end{align}
as desired in \eqref{eq:AandB2}.
We then also have that for $\sum_{i=1}^C \hat X_{ik}$,
\begin{align}
\sum_{i=1}^C \hat X_{ik}\in&\ \left[(1 - \gamma_2)( \sum_{i=1}^C \hat x_{ik} ),\
(1+\gamma_2)(\sum_{i=1}^C \hat x_{ik}  )\right].
\end{align}
Finally, since $\epsilon$ satisfies
$(1 - \gamma_2) / (1 + \gamma_2) \geq 1 - \epsilon$ and 
$(1 + \gamma_2) / (1 - \gamma_2) \leq 1 + \epsilon$
based on our choice of $\gamma_2$,
we have
\begin{align*}
\frac{\hat{X}_{ik}}{\sum_{j=1}^C \hat X_{jk} } \in 
  \left[(1 - \epsilon) 
  \frac{\hat x_{ik}}{\sum_{j=1}^C \hat x_{jk} }
  , 
  (1 + \epsilon) 
  \frac{\hat x_{ik}}{\sum_{j=1}^C \hat x_{jk} }
  \right],
  \text{ for } i \in \{1,\dots,C\}
\end{align*}

This final containment holds whenever the original containment in \eqref{eq:contains} holds,
with probability at least $1-\delta$, completing the proof.
\end{proof}

\subsection{Proof of Lemma 3}
\vspace{2mm}
\noindent{\bf Lemma 3. }{\it
Let $V_1$ and $V_2$ denote a partition of $V = \{1,\dots,n\}$,
and let $R$ be an $n\times n$ matrix
such that for $I \in \{1, 2\}$ and $J \in \{1, 2\}$,
$$\sum_{i\in V_I} R_{ij} = \sum_{i\in V_I} R_{ik} = d_{IJ} 
\text{ for all } j,k\in V_J.
$$ 
Suppose that $\{x_i^k\}_{i=1}^n$ are solutions 
to the recurrence 
\begin{eqnarray}
x_{ik} = \sum_{j=1}^n R_{ij} x_{j,k-1},
\label{eqn:lemc_defx}
\end{eqnarray}
for some specified initial conditions $\{x_{i0}\}_{i=1}^n$,
and that $w_k$, $z_k$ are the solutions 
to the recurrence 
\begin{align}
\begin{pmatrix} w_k \\ z_k \end{pmatrix} 
=
\begin{pmatrix}
d_{11} & d_{12} \\
d_{21} & d_{22}
\end{pmatrix}
\begin{pmatrix} w_{k-1} \\ z_{k-1} \end{pmatrix} 
\label{eqn:lemc_wmult}
\end{align}
with initial conditions
$w_0=\sum_{i\in V_1} x_{i0}$ and $z_0=\sum_{i\in V_2} {x_{i0}}$,
and
where $d_{IJ}$ is as specified above.
Then $w_k = \sum_{i\in V_1} x_{ik}$ and $z_k = \sum_{i\in V_2} x_{ik}$.
}

\begin{proof}
We give a proof by induction in $k$.
First we address the base case, $k=1$.
By expanding the matrix multiplication in \eqref{eqn:lemc_wmult} and isolating
$w_1$ we have:
\begin{align}
w_1 =& d_{11} w_0 + d_{12}z_0\label{eqn:lemc_inducstep1}
 \\
=&d_{11}\sum_{j\in V_1}  x_{j0} +
d_{12} \sum_{j\in V_2} x_{j0} \label{eqn:lemc_inducinitcond}
 \\
=&\sum_{i\in V_1} \sum_{j\in V_1}R_{ij} x_{j0}
+\sum_{i\in V_1}\sum_{j\in V_2} R_{ij} x_{j0} \label{eqn:lemc_inducds}
\\
=&\sum_{i\in V_1} \sum_{j\in V}R_{ij} x_{j0} 
=\sum_{i\in V_1} x_{i1} \label{eqn:lemc_inducdefx}
\end{align}
where in \eqref{eqn:lemc_inducinitcond} we used the initial conditions
$w_0$ and $z_0$, in \eqref{eqn:lemc_inducds} we used the definiton
of $d_{IJ}$, and in \eqref{eqn:lemc_inducdefx} we used the definition
of $x_1$ from \eqref{eqn:lemc_defx}. A similar set of steps can be used to verify the
base case $z_1$.

For our induction we assume that
\begin{align}
w_k =\sum_{i\in V_1}  x_{ik} \text{ and }
z_k =\sum_{i\in V_2}  x_{ik}
\end{align}
and want to show that this implies that
\begin{align}
w_{k+1} =\sum_{i\in V_1}  x_{i,k+1} \text{ and }
z_{k+1} =\sum_{i\in V_2}  x_{i,k+1}.
\end{align}
We follow a similar
set of steps to \eqref{eqn:lemc_inducstep1} - \eqref{eqn:lemc_inducdefx}:
\begin{align}
w_{k+1} &= d_{11} w_k + d_{12}z_k\label{eqn:lemc_inducstep2}
 \\
&=d_{11} \sum_{j\in V_1} x_{jk} +
d_{12} \sum_{j\in V_2} x_{jk} \label{eqn:lemc_induchypothesis}
 \\
&=\sum_{i\in V_1}\sum_{j\in V_1} R_{ij} x_{jk}
+ \sum_{i\in V_1}\sum_{j\in V_2} R_{ij} x_{jk} \label{eqn:lemc_inducds2}
\\
&=\sum_{i\in V_1} \sum_{j\in V}R_{ij} x_{jk} 
=\sum_{i\in V_1} x_{i,k+1} \label{eqn:lemc_inducdefxk}
\end{align}
where we use the induction hypothesis in \eqref{eqn:lemc_induchypothesis},
the definition of $d_{IJ}$ in \eqref{eqn:lemc_inducds2}, and the definition
of $x_{jk}$ from \eqref{eqn:lemc_defx} is used in \eqref{eqn:lemc_inducdefxk}.
Again using a similar set of steps leads to the analogous equality $z_k =
\sum_{i\in V_2} x_{i,k+1}$, thus completing the proof by induction.
\end{proof}

\subsection{Proof of Proposition 2} \text{ }
\\
\noindent{\bf Proposition 2. }{\it
Let $G_n$ be an $n$-node graph generated from a stochastic block model 
$G(n,\pi,P)$ with $C$ communities and $\pi$ fixed, where
$n_i = \pi_i n$ for $i\in\{1,\dots,C\}$, $\sum_{i=1}^C \pi_i=1$, and $p_{ij}>0$ fixed, $\forall i, j$. 
Let $\hat w$ be the geometric discriminant weight vector in the space
of landing probabilities ($1$-step through $K$-step, $K$ fixed)
between two fixed sets of block classes $S$ and $T$ of $G_n$ of size
$n_S = \sum_{i\in S} n_i$ and
$n_T = \sum_{i\in T} n_i$,
where $S$ and $T$ partition $\{1,\dots,C\}$.

 For any $\epsilon, \delta>0$, there exists an $n$ sufficiently large
such that
$
|| n_{S} \hat w - n_{S} \Psi||_1 \le \epsilon
$
with probability at least $ 1-\delta$, where: 

(a) $\Psi = \Psi(\{x_{i,k}\}_{i=1}^C,n,\pi)$ is a vector with coordinates specified by the
solution to a $C$-dimensional linear homogeneous matrix recurrence relation
$x_{ik} = \sum_{j=1}^C n_i p_{ij} x_{j,k-1}$ for $i\in\{1,\dots,C\}$.

(b) If $\sum_{i\in I} n_i p_{ij} =\sum_{i\in I} n_i p_{ik}$
for each $j,k\in J$
for each $I,J\in \{S,T\}$, then $\Psi$ 
is the solution to a $2$-dimensional linear homogeneous matrix recurrence
and the closed-form solution for $\Psi$ follow from
diagonalizing the corresponding $2\times 2$ matrix.

(c) Assuming that the blocks are identically distributed,
then
$$
n_S \Psi_k 
= \left(\dfrac{(p_{in} -p_{out})}{ C p_{out} + (p_{in} -p_{out})}\right)^k,
$$
in which case the geometric discriminant weight vector
is asymptotically equivalent to personalized PageRank.
}

\begin{proof}
By Lemma 2 we have that 
for a fixed walk length $k$,
the landing probabilities $\hat y_{ik}$, $i=1,\ldots,C$,
concentrate around quantities given by the solutions to the $C$-dimensional
linear matrix recurrence $x_{ik} = \sum_{j=1}^C R_{ij} x_{j,k-1}$. 
Specifically, for any
$\epsilon_1>0$, $\delta>0$, with probability at least $1-\delta$:
\begin{align}
n_i \hat y_{ik} \in \left [ (1 - \epsilon_1) \dfrac{x_{ik}}{\sum_{j=1}^{C} x_{jk}},
(1 + \epsilon_1) \dfrac{x_{ik}}{\sum_{j=1}^{C} x_{jk}}
\right ].
\label{eqn:containments2}
\end{align}

We now define $f_k = \sum_{i\in S} x_{ik}$ and $g_k = \sum_{i\in T} x_{ik}$. 
The aggregate probabilities of a $k$-step walk starting at a seed node
in $S$ or $T$ are, respectively,
$\dfrac{\sum_{i\in S} n_i\hat y_{ik}}{n_S}$
and $\dfrac{\sum_{i\in S} n_i\hat y_{ik}}{n_T}$. 
These aggregate probabilities obey the following containment whenever the containments 
\eqref{eqn:containments2} are all satisfied:
\begin{align}
\sum_{i \in S} n_i \hat y_{ik} &\in 
\left [
(1 - \epsilon_1) 
\dfrac{f_k}{f_k+g_k},
(1 + \epsilon_1) 
\dfrac{f_k}{f_k+g_k}
\right ],\\
\sum_{i \in T} n_i \hat y_{ik} &\in
\left [
(1 - \epsilon_1) 
\dfrac{g_k}{f_k+g_k},
(1 + \epsilon_1) 
\dfrac{g_k}{f_k+g_k}
\right ].
\label{eqn:containments3}
\end{align}

The coordinates of the geometric discriminant weight vector 
that we seek to characterize are given by 
\begin{align}
n_S \hat w_k = n_S \left ( 
\dfrac{\sum_{i\in S} n_i \hat y_{ik}}{n_S} - \dfrac{\sum_{i\in T} n_i \hat y_{ik}}{n_T}
\right ).
\label{eqn:prop2_wk}
\end{align}

We obtain the following containments:
\begin{align}
n_S \hat w_k \in
\left [
n_S \Psi_k -  \epsilon_1  \Phi_k
,
n_S  \Psi_k +  \epsilon_1 \Phi_k
\right ],
\hspace{2mm}
\text{ where } 
\hspace{2mm}
\Psi_k = 
\dfrac{1}{f_k+g_k}
\left (
\dfrac{f_k}{n_S} - \dfrac{g_k}{n_T}
\right )
, \hspace{2mm}
\Phi_k =
n_S
\dfrac{
\frac{1}{n_S} f_k + \frac{1}{n_T}g_k
}{
f_k + g_k
}.
\end{align}
If $n_S \le n_T$ then clearly $\Phi_k \le 1$, $\forall k$.
If $n_S > n_T$ then 
$\Phi_k \le n_S/n_T = (\sum_{i \in S} \pi_i)/(\sum_{i \in T} \pi_i)$, $\forall k$.
Since the class proportions $\pi_i$ are all fixed, we can choose a constant 
$C > \max \left [ 1,(\sum_{i \in S} \pi_i)/(\sum_{i \in T} \pi_i) \right ]$ and $\epsilon_1 \le \epsilon/C$ such that:
\begin{align}
|| n_S \hat w - n_S \Psi ||_1 \le  \epsilon,
\end{align}
with probability at least $1-\delta$, as desired.

Notice that $\Psi_k$ depends on both $f_k$ and $g_k$, which in turn depend on all $x_{ik}$, the solutions to a $C$-dimensional recurrence relation. We will now show that when the conditions in part (b) of the proposition are satisfied, the recurrence reduces to a $2$-dimensional recurrence relation in $f_k$ and $g_k$ only.

To compute a closed form expression for $\Psi_k$ we must find a closed form expression for the sums $\sum_{i\in S} x_{ik}$ and $\sum_{i\in T} x_{ik}$. We can now employ Lemma 3,
which requires $\sum_{i\in I}n_i p_{ij} =\sum_{i\in I}n_i p_{ik} $ for each $j,k\in J$,
meaning that 
each node in $I$ has the same expected number of
edges to $J$ for $I,J\in\{S,T\}$. Note that this requirement is less restrictive than the 
two-dimensional stochastic block model, as the partition $S$ and $T$
could have structure: we only require that, for each combination $(I,J) \in \{(S,S),(S,T),(T,S),(T,T)\}$, 
the expected number of edges from a node in $I$ to nodes in $J$
is homogeneous.
This constraint on the degree of nodes within and between 
the partitions indicates that $\sum_{i\in I} R_{ij} = \sum_{i\in I} n_i p_{ij}$ 
satisfies the conditions
required by Lemma 3. Thus to 
compute the closed-form solutions necessary for the geometric
discriminant weight vector
we need not solve the $C$-dimensional matrix recurrence,
but instead can solve a two dimensional recurrence:
\begin{align}
\begin{pmatrix} f_k \\ g_k \end{pmatrix} 
=
\begin{pmatrix}
d_{11} & d_{12} \\
d_{21} & d_{22}
\end{pmatrix}
\begin{pmatrix} f_{k-1} \\ g_{k-1} \end{pmatrix} ,
\label{eqn:lemc_wmult2}
\end{align}
where $f_k=\sum_{i\in S} x_{ik}$
and $g_k=\sum_{i\in T} x_{ik}$ as before,
with initial conditions
$f_0=\sum_{i\in S} x_{i0}$ and $g_0=\sum_{i\in T} x_{i0}$
and
where $d_{IJ} = \sum_{i\in I} n_i p_{ij}$ 
for any choice of $j\in J$
and $I, J\in \{S, T\}$.

Letting $\bf \tilde R = 
\begin{bmatrix}
d_{11} & d_{21} \\
d_{12} & d_{22}
\end{bmatrix}
$ and $\vec C_k=
\begin{bmatrix}
\abar{}{k}\\
\bbar{}{k}
\end{bmatrix} $ 
we have the simple recursion $\vec C_k = \mathbf{\tilde R}\ \vec C_{k-1}$. By induction we
have that $\vec C_k = \mathbf{\tilde R}^k \vec C_0$, where $\vec C_0$ are the initial
conditions. 

\def\twodmat{\mathbf{\tilde R}}

We seek to diagonalize $\twodmat$. When $\twodmat$ is diagonalizable we
have $\twodmat^k =(\textbf{UDU}^{-1}) ^k =  \textbf{UD}^k \textbf{U}^{-1}$, where
$\bf{D}$ is a diagonal matrix with the eigenvalues of $\twodmat$, $\lambda_1$
and $\lambda_2$, along the diagonal, and $\bf{U}$ is a matrix with the
corresponding eigenvectors of $\twodmat$ as its columns. We will derive $\bf U$ and $\bf D$ exactly (below) and
thus can derive $\vec C_k$ exactly for all $k$.
In this case,
we denote the eigenvalues of $\twodmat$ and the matrix containing
its eigenvectors are given by
$\lambda_1, \lambda_2$, $\mathbf{U}$:
\begin{align}
\lambda_{1} =
\dfrac{1}{2}\left(d_{11} + d_{22} -
\phi\right), \ 
\lambda_{2} =
\dfrac{1}{2}\left(d_{11} + d_{22} +
\phi\right),
\
\textbf{U} = 
\begin{pmatrix}
\dfrac{d_{11} - d_{22} - \phi} {2d_{12}} &
\dfrac{d_{11} - d_{22} + \phi}{2d_{12}} 
\\[2ex]
1 & 1
\end{pmatrix},
\label{eqn:phi}
\end{align}
where $\phi = \sqrt{(d_{11} - d_{22})^2 + 4 d_{12} d_{21}}$.

To establish the above solutions to the diagonalization
we will show that 
$\twodmat \textbf{U}= \textbf{U}  \textbf{D}$
for $\bf{U}$, $\bf D$ specified
in
\eqref{eqn:phi}. We begin by respectively multiplying the two pairs of matrices:
\begin{align}
\twodmat \textbf{U}= &
\begin{pmatrix}
\dfrac{d_{11}}{2 d_{12}} (d_{11} - d_{22} - \phi) + d_{21} &
\dfrac{d_{11}}{2 d_{12}} (d_{11} - d_{22} + \phi) + d_{21} \\[2ex]
\dfrac{1}{2} (d_{11} - d_{22} - \phi) + d_{22} &
\dfrac{1}{2} (d_{11} - d_{22} + \phi) + d_{22}
\end{pmatrix},\\
\textbf{U} \textbf{D}= &
\begin{pmatrix}
\dfrac{\lambda_1(d_{11} - d_{22} - \phi)} {2d_{12}} &
\dfrac{\lambda_2(d_{11} - d_{22} + \phi)}{2d_{12}} 
\\[2ex]
\lambda_1 & \lambda_2
\end{pmatrix},
\end{align}
and immediately can observe that indeed $(\twodmat\mathbf{U})_{12} = (\textbf{UD})_{12}$ 
and $(\twodmat\mathbf{U})_{22} = (\textbf{UD})_{22}$. 
By distributing $\lambda_1$ and $\lambda_2$ in
$(\textbf{UD})_{11}$ and $(\textbf{UD})_{21}$ and simplifying
we can observe:
\begin{align}
(\twodmat\textbf{U})_{11}  
- (\textbf{UD})_{11} = 
(\twodmat\textbf{U})_{21} 
- (\textbf{UD})_{21} 
= \dfrac{(d_{11} - d_{22})^2 + 4 d_{12} d_{21} - \phi^2}{4 d_{12}} = 0
\end{align}
where in the last step we use the definition of $\phi$ as $\phi=\sqrt{(d_{11} - d_{22})^2 + 4 d_{12} d_{21}}$. We thus 
establishe that $\twodmat\textbf{U} = \textbf{UD}$. 

The recurrence in \eqref{eqn:lemc_wmult2} has the following general solution in terms of the eigenvectors
and eigenvalues of $\twodmat$:
\begin{align}
\abar{}{k} = &
\dfrac{\lambda_1^k U_{11} (\bbar{}{0} U_{12} - \abar{}{}_0 U_{22}) 
- 
\lambda_{2}^k U_{12} (\bbar{}{}_0 U_{11} - \abar{}{0} U_{21})}
{U_{12} U_{21} - U_{11} U_{22}}
\label{eqn:genrecur1}\\
\bbar{}{k} = &
\dfrac{\lambda_1^k U_{21} (\bbar{}{}_0 U_{12} - \abar{}{}_0 U_{22}) 
- 
\lambda_2^k U_{22} (\bbar{}{0} U_{11} - \abar{}{0} U_{21})}
{U_{12} U_{21} - U_{11} U_{22}}.
\label{eqn:genrecur2}
\end{align}
Assuming $d_{11} + d_{12} = d_{21} + d_{22}$, as assumed
in part (b) of the proposition statement,
leads to 
$\lambda_1 = d_{12} + d_{22}$, $\lambda_2 = -d_{21} + d_{22}$,
and 
$\mathbf{U}~=~\begin{pmatrix}
d_{12}/d_{21}& -1\\ 1&1
\end{pmatrix}$.
Thus \eqref{eqn:genrecur1} and \eqref{eqn:genrecur2} simplify to
\begin{align}
f_k=\dfrac{d_{12} \lambda_1^k + d_{21} \lambda_2^k}{d_{12} + d_{21}},\
g_k=\dfrac{d_{21} (\lambda_1^k -        \lambda_2^k)}{d_{12} + d_{21}}.
\end{align}
We thus establish part (b) of the proposition, that $\Psi_k$ depends only on the solution of a $2$-dimensional recurrence relation.

For the special case of the blocks being 
characterized by identical
parameters, $n_S \Psi_k$ reduces to
weights equivalent to personalized PageRank for a particular choice of $\alpha$.
To see this, we will denote the number of nodes in each of the
$C$ identically distributed blocks as $N = n / C$,
and the number of blocks in the in- and out-classes, $S$ and $T$, as $|S|$ and
$|T|$, where $|S|$ and $|T|$ are integers that sum to $C$. 

Thus in this case we have that
\begin{align}
d_{11} = N ((|S| - 1) p_{out} + p_{in}),\  d_{12} = N |S| p_{out},\notag\\
d_{22} = N ((|T| - 1) p_{out} + p_{in}),\  d_{21} = N |T| p_{out}, 
\label{eqn:dspecial}
\end{align}
where $\lambda_1$, $\lambda_2$, and $f_k + g_k$ reduce to:
\begin{align}
\lambda_1& = \dfrac{1}{2}\left( N|S| p_{out} + N((|T| - 1 ) p_{out} + p_{in}) \right) 
 =  n p_{out} + \frac{n}{C}(p_{in} -p_{out}), \\
\lambda_2 &= -N|T| p_{out} + N((|T| - 1 ) p_{out} + p_{in})
 = \frac{n}{C}(p_{in}- p_{out}),\\
f_k &= \dfrac{|S| \lambda_1^k + |T| \lambda_2^k}{C},
\\
g_k &= \dfrac{|T|(\lambda_1^k -\lambda_2^k)}{C},
\end{align}
and $n_S \Psi_k$ reduces to:
\begin{align}
n_S \Psi_k = &
\dfrac{n_S}{f_k+g_k} \left(\dfrac{f_k}{n_S} - 
\dfrac{g_k}{n_T}\right) 
=
\dfrac{n_S}{n} \dfrac{1}{\lambda_1^k}  \left(
\left(\lambda_1^k + \dfrac{|T|}{|S|}\lambda_2^k\right)
- \left(\lambda_1^k - \lambda_2^k\right)
\right)
=
 \dfrac{|S|}{C} \left(\dfrac{|T|}{|S|}+1\right)\left(\dfrac{\lambda_2}{\lambda_1}\right)^k \\
=&\left(\dfrac{(p_{in} -p_{out})}{ C p_{out} + (p_{in} -p_{out})}\right)^k.
\label{eqn:psisimple}
\end{align}
The weights derived here are again precisely the weights employed by personalized PageRank for
$\alpha=\dfrac{p_{in} -p_{out}}{C p_{out} + (p_{in} -p_{out})}$.
\end{proof}
We see that Proposition 1 is simply a special case of Proposition 2: for $C=2$ we recover $\alpha = \dfrac{p_{in} -p_{out}}{p_{in} + p_{out}}$, as in Proposition 1.

\section{Details of Belief Propagation}

\def\pij{p_{q_i,q_j}}
\def\qs{\{q_i \}_{i=1}^n}

Given a $C$-block stochastic block model $G(n,\pi,P)$, the probability of observing an adjacency matrix $A$ and block assignments $\{ q_i \}_{i=1}^n$ --- where $q_i \in \{1,\ldots,C\}$ assigns node $i$ to a block class --- with parameters $\theta = (C, \pi, P)$ is:
\begin{eqnarray}
\Pr (A, \qs | \theta ) = 
\prod_{\substack {i,j=1 \\ j \ne i}}^n 
\pij^{A_{ij}}
(1-\pij)^{1-A_{ij}}
\prod_{k=1}^C
\pi_{q_{k}}.
\end{eqnarray}
When an adjacency matrix $A$ has been observed, we obtain the following log-likelihood function for $\qs$:
\begin{eqnarray}
\ell(\qs; A, \theta) = 
\sum_{\substack {i,j=1 \\ j \ne i}}^n 
\bigg (
A_{ij} \log \pij
+ (1- A_{ij}) \log (1-\pij)
\bigg )
+ \sum_{k=1}^n \log \pi_{q_k}.
\end{eqnarray}
Maximizing this log-likelihood over possible assignments $\qs$ is NP-hard in general (by a reduction from graph bisection), though an EM algorithm can be applied to find local maxima \cite{snijders1997estimation}. The EM approach is understood to perform decently well in graphs where there is sufficiently unambiguous block structure, but recent work has shown that the more powerful method known as belief propagation (BP) is able to recover labels that are correlated with the ground truth labels under considerably weaker conditions on the unambiguousness of the block structure \cite{decelle2011asymptotic}. 

Belief propagation (BP) has been found to make much better inferences than EM
approaches in practice, reaching the known {\it resolution limit} (also called the {\it detectability threshold}) for the
problem \cite{fortunato2007resolution}. We describe belief propagation here in order to benchmark our
classification algorithms against the best known methods for the problem. It is
important to keep in mind that belief propagation is a much more empowered
and computationally demanding 
algorithm than the classification algorithms Lin-SBMRank and Quad-SBMRank that we
present and study in the main text of this work. Those algorithms 
are all restricted to simple discriminative classification rules in the
space of random walk landing probabilities, classifying individual nodes
independently based on these probabilities. In contrast, belief propagation
performs a global joint inference of all node labels. 

While BP lacks rigorous guarantees for general stochastic block models, two recent algorithms \cite{mossel2013proof,massoulie2014community} inspired by the basic mechanics of BP have recently admitted rigorous analyses on sparse graphs, showing that they are capable of identifying labels correlated with the true labels all the way up to the known resolution limit for the problem. Both algorithms are acknowledged as impractical in practice and no implementation has been studied. A recent semidefinite programming relaxation of the likelihood maximization should also be of interest to the reader \cite{abbe2016exact}; it has been implemented in practice and found to be quite effective, though its analysis obtains slightly weaker rigorous guarantees than the rigorous analyses of the two other recent algorithms above. We do not employ any of these three algorithms, 
focusing instead on ordinary belief propagation.

The analysis of belief propagation tends to rely heavily on metaphors from statistical physics. In this section we outline the basic motivation for belief propagation's performance, and then specify the precise instantiation of belief propagation that we employ in this work.

\xhdr{Belief Propagation and SBMs in theory}
Belief propagation (BP) is a message passing algorithm for the inference of joint distributions of random variables with conditional dependencies represented by graphs. Belief propagation infers marginal distributions of 
unobserved variables of such models, commonly called {\it graphical models}. 
When the graph underlying a graphical model is a tree,
the BP algorithm is known to converge on a fixed point that minimizes an objective function known as the {\it Bethe free energy} of the joint probability distribution represented by the model  \cite{yedidia2003understanding}. On more general graphs, when it does converge then the fixed point it converges upon must be a stationary point (though not necessarily a global minima) of the Bethe free energy. Belief propagation is therefore widely applied beyond the context of trees, in what is sometimes specified as {\it loopy belief propagation}, named after the presence of cycles (loops) in the graph. The procedure we describe here is an instance of loopy belief propagation, though we refer to it simply as belief propagation (or BP).

Belief propagation has recently been adapted for the inference of block labels for graphs realized from stochastic block models, where the block labels are viewed as unobserved latent variables \cite{decelle2011asymptotic}. 
It can be shown that any minimum of the Bethe free energy is a maximum of the log-likelihood of the stochastic block model  \cite{neal1998view,zhang2012comparative}.
Thus, if a graph realized under the SBM is a tree then belief propagation will quickly converge upon the maximum likelihood estimate of the SBM parameters (these parameters include the assignment of nodes to block classes). Note that this does not contradict the NP-hardness of maximizing the likelihood on graphs in general.
For a graph realized under the SBM that is not a tree, the convergence upon the global maximum of the likelihood under BP is no longer guaranteed. In fact, loopy BP is not guaranteed to converge on any solution at all, though some sufficient conditions for convergence on non-trees are known \cite{mooij2007sufficient}. When it does converge, however, it is a stationary point of the likelihood. 

Despite the lack of rigorous results, BP is widely understood to find good solutions (solutions with near-maximal likelihood) in practice.  We study BP for blocks of equal size only, however note that recent work have successfully adapted BP to block models with unequal group sizes \cite{zhang2016community}.

A method known as {\it belief optimization} is worth mentioning, as it attempts to optimize the Bethe free energy directly to a local minima, thus avoiding BPs lack of convergence guarantees \cite{welling2001belief}. This approach is closely related to maximizing the stochastic block model likelihood directly via EM \cite{snijders1997estimation}. Belief optimization is slower than BP when BP does converge, and thus is less often used in practice.  

\xhdr{Belief Propagation and SBMs in practice}
We now present a distilled presentation of sparse-SBM-BP, the instantiation of belief propagation that we employ for inferring latent node labels in stochastic block models. The presentation here is largely a reproduction of the derivation in \cite{decelle2011asymptotic} with added clarifications. Our derivations are specific to the case of $C$ blocks with equal sizes $n$, forming a graph on $N=Cn$ nodes. Let $c_{sr}=n p_{sr}$ be the known expected degree between class $s$ and $r$ and let $\pi_s=C/n$ denote the proportion of nodes in block $s$.

Let $\psi_s^i$ denote the {\it belief} that node $i$ belongs to class $s$. 
The BP equations are then defined as the following recursively defined conditional marginal probabilities $\psi_{s}^{i \rightarrow j}$, also called {\it messages}:
\begin{eqnarray}
\label{eqn:bp1}
\psi_{s}^{i \rightarrow j} = \frac{1}{Z^{i \rightarrow j}} \pi_{s} 
\prod_{\substack{k=1 \\ k \ne j}}^n 
\bigg [ \sum_{r=1}^C c_{s,r}^{A_{ik}} 
\big (1-\frac{c_{s,r}}{N} \big )^{1-A_{ik}}
\psi_{r}^{k \rightarrow i}
\bigg ], \forall i,j,s,
\end{eqnarray}
where the beliefs $\psi_t^i$ are aggregated up as:
\begin{eqnarray}
\label{eqn:bp2}
\psi_{s}^i = \frac{1}{Z^i} \pi_{s} 
\prod_{j=1}^n 
\bigg [ \sum_{r=1}^C c_{s,r}^{A_{ij}} 
\big (1-\frac{c_{s,r}}{N} \big )^{1-A_{ij}}
\psi_{r}^{j \rightarrow i}
\bigg ], \forall i,s.
\end{eqnarray}
In these two equations $Z^{i \rightarrow j}$ and $Z^i$ are both normalizing constants determined by the constraints $\sum_{t=1}^C \psi_t^{i \rightarrow j} = 1$ and $\sum_{t=1}^C \psi_t^i = 1$.

The BP algorithm consists of iterating the BP equations in \eqref{eqn:bp1} until, hopefully, convergence. When these BP equations converge to a fixed point then the beliefs $\psi_{s}^i$ in \eqref{eqn:bp2} are a stationary point of the Bethe free energy of the graphical model underlying the stochastic block model, an objective with close ties to the model likelihood as discussed above.

Notice that each of the $O(N^2)$ message equations in \eqref{eqn:bp1} contains $N-1$ terms, meaning that each iteration of the BP equations requires $O(N^3)$ computation. For sparse graphs the following heuristic approximation dramatically simplifies the computation \cite{decelle2011asymptotic}. We first separate the product in \eqref{eqn:bp1} into neighbor nodes ($k \in \partial i$) and non-neighbor nodes ($k \notin \partial i$):
\begin{eqnarray}
\psi_{s}^{i \rightarrow j} = \frac{1}{Z^{i \rightarrow j}} \pi_{s} 
\prod_{\substack{k \notin \partial i  \\ k \ne j}} 
\bigg [ 1 - \frac{1}{N} \sum_{r=1}^C
c_{s,r}
\psi_{r}^{k \rightarrow i}
\bigg ]
\prod_{\substack{k \in \partial i \\ k \ne j}} 
\bigg [ \sum_{r=1}^C 
c_{s,r} 
\psi_{r}^{k \rightarrow i}
\bigg ]
, \forall i,j,s.
\end{eqnarray}

We now employ a sequence of three heuristic approximations. First, we use that $(1-x) \approx \exp(-x)$ for small $x$:
\begin{eqnarray}
\psi_{s}^{i \rightarrow j} = \frac{1}{Z^{i \rightarrow j}} \pi_{s} 
\exp \left (  - \frac{1}{N}
\sum_{\substack{k \notin \partial i \\ k \ne j}} 
 \sum_{r=1}^C
c_{s,r}
\psi_{r}^{k \rightarrow i}
\right )
\prod_{\substack{k \in \partial i \\ k \ne j}} 
\bigg [ \sum_{r=1}^C 
c_{s,r} 
\psi_{r}^{k \rightarrow i}
\bigg ]
, \forall i,j,s.
\end{eqnarray}
Second, for non-neighbors in the internal sum the difference between summing over strictly non-neighbors $k \notin \partial i$ and all nodes $k=1,...,N$ is small:
\begin{eqnarray}
\psi_{s}^{i \rightarrow j} = \frac{1}{Z^{i \rightarrow j}} \pi_{s} 
\exp \left ( - \frac{1}{N}
\sum_{k=1}^N 
 \sum_{r=1}^C
c_{s,r}
\psi_{r}^{k \rightarrow i}
\right )
\prod_{\substack{k \in \partial i \\ k \ne j}} 
\bigg [ \sum_{r=1}^C 
c_{s,r} 
\psi_{r}^{k \rightarrow i}
\bigg ]
, \forall i,j,s.
\end{eqnarray}
Third, for non-neighbor pairs the messages $\psi_{r}^{k \rightarrow i}$ are approximately the beliefs $\psi_{r}^{k}$, $\psi_{r}^{k \rightarrow i} = \psi_{r}^{k} + O(1/N)$, for all $r$ and all $k$ independent of the destination $i$. We obtain:
\begin{eqnarray}
\psi_{s}^{i \rightarrow j} = \frac{1}{Z^{i \rightarrow j}} \pi_{s} 
\exp \left ( - \frac{1}{N}
\sum_{k=1}^N 
 \sum_{r=1}^C
c_{s,r}
\psi_{r}^{k}
\right )
\prod_{\substack{k \in \partial i}} 
\bigg [ \sum_{r=1}^C 
c_{s,r} 
\psi_{r}^{k \rightarrow i}
\bigg ]
, \forall i,j,s.
\end{eqnarray}
We now observe that the terms $\xi_s := \pi_s \exp \left ( - \frac{1}{N}
\sum_{k=1}^N 
 \sum_{r=1}^C
c_{s,r}
\psi_{r}^{k}
\right )$
do not depend on $i$ or $j$ and can be computed for each $s$ once per iteration of the BP equations. 
We further note that the only message equations used in the iteration are those for $(i,j)$ pairs
that are neighbors in the graph. We thus obtain the following simplified {\it sparse BP} equations for stochastic block model label inference, replacing the dense equations in \eqref{eqn:bp1} and \eqref{eqn:bp2} with:

\begin{eqnarray}
\label{eqn:sparsebp1}
\psi_{s}^{i \rightarrow j} &=& \frac{1}{Z^{i \rightarrow j}} 
\xi_s
\prod_{\substack{k \in \partial i \\ k \ne j }} 
\bigg [ \sum_{r=1}^C 
c_{s,r} 
\psi_{r}^{k \rightarrow i}
\bigg ]
, \hspace{1.1cm} (i,j)\in E, \text{ } s=1,...,C, 
\\
\label{eqn:sparsebp2}
\xi_s &=& \pi_s \exp \left ( - \frac{1}{N}
\sum_{k=1}^N 
 \sum_{r=1}^C
c_{s,r}
\psi_{r}^{k}
\right ), \hspace{1cm} s=1,...,C,
\\
\label{eqn:sparsebp3}
\psi_{s}^i &=& \frac{1}{Z^i} 
\xi_s
\prod_{\substack{j \in \partial i }} 
\bigg [ \sum_{r=1}^C 
c_{s,r} 
\psi_{r}^{j \rightarrow i}
\bigg ]
, \hspace{1.6cm} i=1,...,N,\text{ } s=1,...,C.
\end{eqnarray}
Here, as before,  $Z^{i \rightarrow j}$ and $Z^i$ are normalizing constants determined by the constraints $\sum_{t=1}^C \psi_t^{i \rightarrow j} = 1$ and $\sum_{t=1}^C \psi_t^i = 1$. For a sparse graph where the maximum degree $d$ is bounded $O(1)$ in $n$ there are just $O(dN)$ BP equations with an overall computational cost is thus $O(d^2 N)$ per iteration, much less than $O(N^3)$ for the standard BP algorithm before the heuristics were applied.

Notice that in the original BP equations \eqref{eqn:bp1} only the messages $\psi_{s}^{i \rightarrow j}$ were computed each iteration, whereas the sparse BP equations in Eqs.~(\ref{eqn:sparsebp1})-(\ref{eqn:sparsebp3}) above compute the beliefs $\psi_{s}^{i}$ as part of computing the messages $\psi_{s}^{i \rightarrow j}$. Once the messages converge upon a fixed point then the beliefs can be used for classification by assigning nodes the class with the highest belief.

\xhdr{Initial conditions} 
The sparse BP equations in Eqs.~(\ref{eqn:sparsebp1})-(\ref{eqn:sparsebp3}) are defined recursively, and must begin somewhere. In this work we employ the following initializations. We begin by setting the beliefs according to the known class proportions, $\psi_{s}^{i,(0)} = \pi_s$ for $s=1,...,C$. For the case of $\pi_s=1/C$ (balanced class sizes) we then obtain initial conditions on each $\xi_s$: 
\begin{eqnarray}
\xi_s^{(0)} &=& \frac{1}{C} \exp \left ( 
- \frac{1}{NC}
\sum_{r=1}^C
c_{s,r}
\right ), \hspace{1.6cm} s=1,...,C.
\end{eqnarray}
Lastly, we initialize the messages $\psi_{s}^{i \rightarrow j,(0)}$ uniformly at random. As we require that $\sum_{s=1}^C \psi_{s}^{i \rightarrow j}=1$, this uniformly random initialization is achieved by either (a) sampling each message independently from the uniform distribution on $[0,1]$ and then normalizing each message by the sum over all messages corresponding to their $(i,j)$ pair,
or equivalently (b) sampling each vector of messages corresponding to an $(i,j)$ pair from Dir($1,\ldots,1$), the uniform Dirichlet distribution.

We halt the iteration when the equations in \eqref{eqn:sparsebp1} are converged below some predefined numerical 
tolerance on the update difference or the number of iterations exceeds some predefined max iteration count.

\xhdr{Application to the seed set expansion problem}
Belief propagation performs an optimization over all labelings of the nodes of the graph, without necessarily utilizing any information about labelled nodes.  In order to adapt BP to the seed set expansion problem that we study, we simply fix the beliefs about the seed set such that if the seed set $S \subset V$ belongs to class $s$ then:
\begin{eqnarray}
\psi_{s}^i = 1, \forall i \in S, \hspace{1cm}
\psi_{t}^i = 0, \forall i \in S, t \ne s.
\end{eqnarray}
These beliefs are given at initialization and also maintained during the iteration of the BP equations; the beliefs about the seed set are not updated. Beliefs about all other nodes $i \notin S$ (those nodes not in the labeled seed set) are initialized according to the known class proportions minus an appropriate adjustment for what is known about the seed set:
\begin{eqnarray}
\psi_{s}^i = \frac{\frac{1}{C} n - |S|}{n-|S|}, \forall i \notin S,  \hspace{1cm}
\psi_{t}^i = \frac{\frac{1}{C} n}{n-|S|}, \forall i \notin S, t \ne s.
\end{eqnarray}

\section{Additional Computational Results and Discussion}

\xhdr{Exemplar SBM Parameters}
In Figure 1A of the main text we illustrate
the adjacency matrix for a four-block SBM
with $n = 2048$, 
$\pi = [491/2048, 532/2048, 471/2048, 554/2048]$
and 
$ P =
\left(
\begin{smallmatrix}
0.4       &  0.15      &  0.08      &  0.04      \\
0.15&  0.38      &  0.04&  0.08\\
0.06      &  0.08&  0.37      &  0.16\\
0.06&  0.04&  0.18&  0.36      \end{smallmatrix}
\right)
$.
If considering a seed node from block 1, the block partition $S = \{1, 2\}$ 
and $T=\{3, 4\}$ leads this 4-block SBM with non-identical blocks 
to satisfy the portions of Proposition 2 
that apply to non-identical block models.

\xhdr{Numerical Considerations for Covariance Matrices}
The covariance matrices describe the covariances
between the landing probabilities for a random
walk starting at the seed node and walking $1$ to
$K$ steps. For large step counts the
landing probabilities begin to converge upon 
the stationary distribution of a random walk, 
meaning that the covariance
between step $K-1$ and $K$ becomes very high.
In general the last several columns of the
covariances matrices become strongly collinear
for large values of $K$. 
To mitigate against ill-conditioned matrix
inversions, we restrict the maximum number of
steps $K$ included in our landing probability space
in a manner that keeps the condition numbers $\kappa(\Sigma_a)$ and 
$\kappa(\Sigma_b)$ both below 
$10^{10}$. In practice this empirically amounts to performing
our analysis in the space of landing probabilities for the first $K=6$ steps.

\begin{figure*}[t]
\centering
\includegraphics[width=0.25\textwidth]{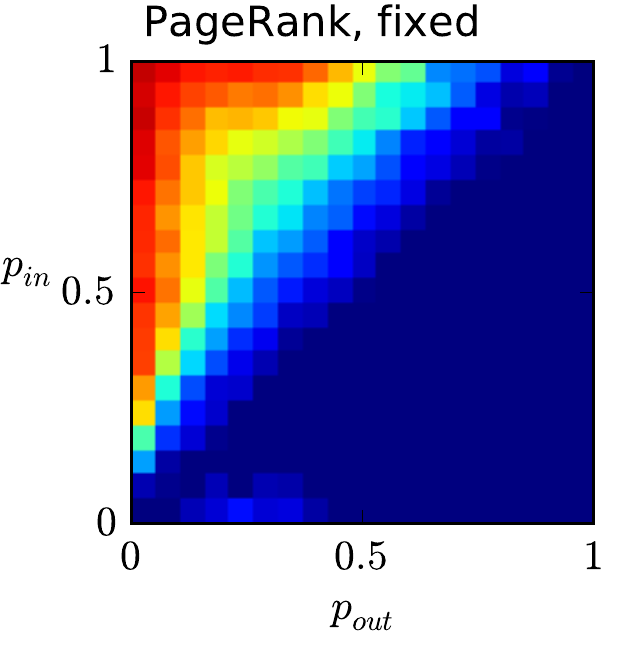}
\includegraphics[width=0.25\textwidth]{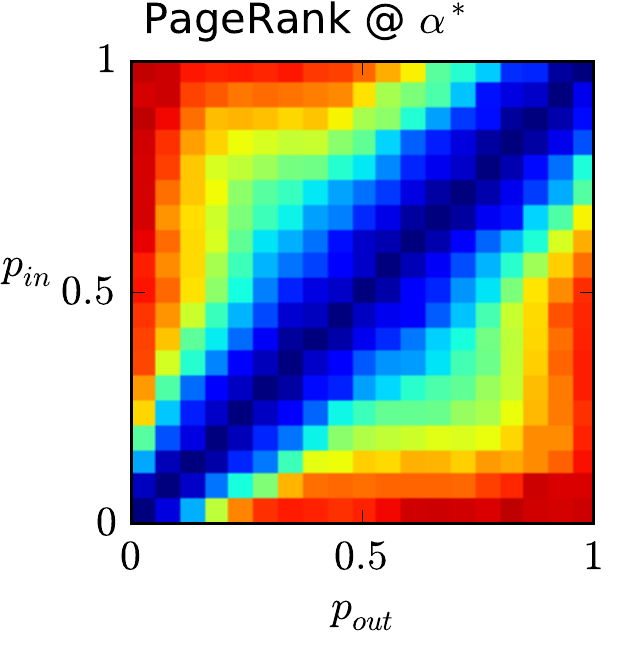}
\includegraphics[width=0.25\textwidth]{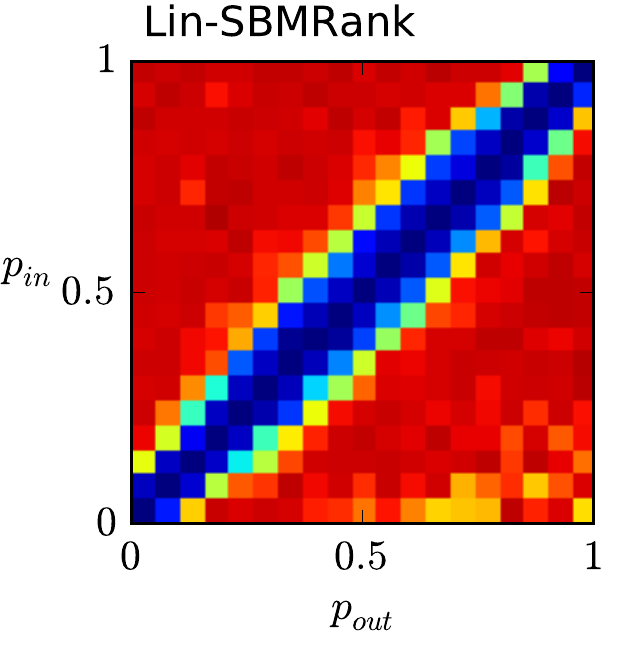}
\caption{
(A)-(C) Heatmaps: the Pearson correlation $r$ between the inferred partition and true partition as a function of $p_{aa}=p_{bb}=p_{in}$ and $p_{ab}=p_{out}$ in four settings.
Red is high ($r=1$) and blue is low ($r=0$).
Using a personalized PageRank discriminant function ranking based on (A) a fixed choice of
$\alpha=0.7$ (corresponding to $p_{in}=0.6$, $p_{out}=0.1$), (B)
$\alpha^*$ set by the true values of $p_{in}$ and $p_{out}$, and (C) our
normalized linear function Lin-SBMRank, $g_2(r) = \Sigma^{-1}(a-b)^Tr$, for the true values of $p_{in}$ and $p_{out}$. The
quadratic discriminant function Quad-SBMRank, not shown, is similar
to Lin-SBMRank.
}
\label{f:sbm_resultsSI}
\end{figure*}

\xhdr{Further Performance Results For SBMs}
As an illustration of the improved performance
of our algorithm for recovering partitions correlated
with the ground truth, Figure~\ref{f:sbm_resultsSI}
shows heat maps of the Pearson correlation
$r$ of various methods as a function of $p_{in}$ and $p_{out}$.
We clearly see that our normalized linear classification
performs significantly better through the space of
stochastic block model parameters;
the quadratic classification (not shown in this figure)
produces a heat map that's visually very similar to the one for
the normalized linear classification (with slightly higher performance).
We note that all these methods require knowledge or estimation
of the underlying parameters --- in Figure~\ref{f:sbm_resultsSI}A
we see that when the discriminant function is configured
with a fixed set of parameters that are far from the
true values the classification can be quite poor.

\xhdr{SBM Parameter Estimation}
We employ the following consistent estimators for the parameters of a stochastic
block model $G((n_a, n_b), P)$ where $p_{11} = p_{22} = p_{in}$, $p_{12} = p_{out}$,
also known as the affiliation model, 
as due to Allman et al. \cite{allman2011parameter}. 
Given an observed adjacency matrix $A$, then
\begin{align}
\hat p_{out} =
\frac{
(s_3 - s_2 s_3) m_1^3 + (s_2^3-s_3)m_2 m_1 + (s_3 s_2 - s_2^3)m_3
}{
(m_1^2 - m_2)(2s_2^3 - 3s_3s_2 + s_3)
}, \hspace{1cm}
\hat p_{in} =
\frac{
m_1 + (s_2 -1) \hat p_{out}
}{
s_2
},
\end{align}
where
\begin{align}
s_2 &= n_a^2 + n_b^2,\\
s_3 &= n_a^3 + n_b^3,\\
m_1 &= \frac{1}{n(n-1)} \sum_{i,j=1, i \ne j}^n A_{ij},\\
m_2 &= \frac{1}{n(n-1)(n-2)} \sum_{i,j,k=1, i \ne j \ne k }^n A_{ij} A_{ik},\\
m_3 &= \frac{1}{n(n-1)(n-2)} \sum_{i,j,k=1, i \ne j \ne k }^n A_{ij} A_{ik} A_{jk}.
\end{align}

\xhdr{Acknowledgements.}
Supported in part by a Simons Investigator Award, an ARO MURI grant,
a Google Research Grant, a Facebook Faculty Research Grant, and a David
Morgenthaler II Faculty Fellowship.

\bibliographystyle{plain}
\bibliography{blockrefs}

\end{document}